\documentclass[journal,comsoc]{IEEEtran}
\usepackage[dvips]{graphicx}
\usepackage{latexsym}
\usepackage{epsfig}
\usepackage{color}
\usepackage{amsmath} 
\usepackage{amssymb}
\usepackage{amsxtra}
\usepackage{enumitem}

\usepackage{longtable}

\usepackage[T1]{fontenc}
\usepackage{setspace}
\usepackage{array,multirow}

\usepackage[cmintegrals]{newtxmath}
\usepackage{graphicx}
\graphicspath{{Figures/}}
\usepackage{url}
\usepackage{cite}
\usepackage{algorithm}

\usepackage{algpseudocode}

\makeatletter
\def\BState{\State\hskip-\ALG@thistlm}
\makeatother

\usepackage{epstopdf}
\usepackage{amsxtra}
\usepackage{amstext}
\usepackage{amssymb}
\usepackage{latexsym}
\usepackage{dsfont} 
\usepackage{color}
\usepackage{multirow}
\usepackage{float}
\usepackage{hyperref}

\usepackage{tabularx,colortbl}
\usepackage[table]{xcolor}

\usepackage{lscape}

\usepackage{algorithm}
\usepackage{algpseudocode}

\usepackage{units}
\usepackage{cases}
\usepackage[ subrefformat=parens,labelformat=parens, caption=false, font=footnotesize]{subfig}
\usepackage{mathtools}

\newcommand{\mbf}[1]{\mathbf{#1}}

\newcommand{\nth}[1]{{#1}{\text{th}}}

\newcommand{\abs}[1]{\left|{#1}\right|}
\newcommand{\norm}[1]{\left\|{#1}\right\|}

\newcommand{\Prb}{\mathrm{Pr}}

\begin{document}

\title{Signal Processing and Machine Learning Techniques for Terahertz Sensing: An Overview}

\author{Sara~Helal,
        Hadi~Sarieddeen,~\IEEEmembership{Member,~IEEE,}
        Hayssam~Dahrouj,~\IEEEmembership{Senior Member,~IEEE,}
        Tareq~Y.~Al-Naffouri,~\IEEEmembership{Senior Member,~IEEE,}
        and~Mohamed-Slim~Alouini,~\IEEEmembership{Fellow,~IEEE}
\thanks{S. Helal is with the Department of Electrical Engineering, Effat University, Jeddah 22332, Saudi Arabia (e-mail:sahelal@effat.edu.sa).

H. Sarieddeen, T. Y. Al-Naffouri, and M.-S. Alouini are with the Division of Computer, Electrical and Mathematical Sciences and Engineering, King Abdullah University of Science and Technology, Thuwal 23955-6900, Saudi Arabia (e-mail: hadi.sarieddeen@kaust.edu.sa; tareq.alnaffouri@kaust.edu.sa; slim.alouini@kaust.edu.sa).

H. Dahrouj is with the Center of Excellence for NEOM Research, King Abdullah University of Science and Technology,  Thuwal 23955-6900, Saudi Arabia (e-mail: hayssam.dahrouj@gmail.com).

This work was supported, in part, by the King Abdullah University of Science and Technology (KAUST) Office of Sponsored Research (OSR) under Award ORA-CRG2021-4695, and Center of Excellence for NEOM Research.
}

}

\maketitle

\begin{abstract}

Following the recent progress in Terahertz (THz) signal generation and radiation methods, joint THz communications and sensing applications are being proposed for future wireless systems. Towards this end, THz spectroscopy is expected to be carried over user equipment devices to identify material and gaseous components of interest. THz-specific signal processing techniques should complement this re-surged interest in THz sensing for efficient utilization of the THz band. In this paper, we present an overview of these techniques, with an emphasis on signal pre-processing (standard normal variate normalization, min-max normalization, and Savitzky-Golay filtering), feature extraction (principal component analysis, partial least squares, t-distributed stochastic neighbor embedding, and nonnegative matrix factorization), and classification techniques (support vector machines, k-nearest neighbor, discriminant analysis, and naive Bayes). We also address the effectiveness of deep learning techniques by exploring their promising sensing and localization capabilities at the THz band. Lastly, we investigate the performance and complexity trade-offs of the studied methods in the context of joint communications and sensing (JCAS). We thereby motivate the corresponding use-cases, and present a handful of contextual future research directions.

\end{abstract}

\IEEEpeerreviewmaketitle

\section{Introduction} 

\subsection{History and motivation for THz sensing}

The introduction of Terahertz (THz) technology has boosted research on exploring the atomic behavior of materials, fostering diverse opportunities in wireless sensing \cite{bogue2018sensing} and imaging \cite{articleJepsen}. The THz band, which extends from 100 Gigahertz (GHz) to 10 THz (wavelength from $\sim$ \unit[1]{mm} to \unit[0.1]{mm}) is sandwiched between the well-studied microwave and infrared bands of the electromagnetic spectrum. This far-infrared region is thus seen as a transition region where optics meets electronics. In the optical domain, THz radiation, also known as T-ray, is treated as a beam of light that can be manipulated by mirrors and lenses, of which the intensity of light can be measured. On the other hand, in electronics, T-ray is treated as an electrical wave, of which the phase of the electric field can be measured. Due to the lack of efficient and reliable sources, the THz band has been perceived as the THz gap. However, the development of high-energy strong-field THz generation and detection devices is bridging this gap. In particular, the THz generation and detection technologies \cite{sengupta2018terahertz} are classified as optical, electronic, hybrid photonic-electronic, and plasmonic (graphene-based).

THz wireless sensing leverages the inherent spectral fingerprints of molecules at the THz band for various sensing and imaging applications, in the areas of quality control, food safety, security, health, astronomy, and environmental monitoring. Owing to its low photon energy ($\unit[4]{meV}$ at $\unit[1]{THz}$), T-ray is non-ionizing; therefore, it can analyze the content of biological bodies without posing significant harm. Many biological and chemical materials exhibit unique spectral fingerprint information in the THz range. For instance, THz vibrational modes enable the study of molecular structures and vibrational dynamics such as scissoring, wagging, and twisting that arise from both intermolecular and intramolecular interactions. THz radiation also has special interactions and strong molecular coupling with hydrogen-bonded networks, which maximizes the potential of observing water dynamics. Furthermore, T-rays can penetrate various non-conducting, amorphous, and dielectric materials and are highly reflected from metals, which gives T-rays the capability to inspect metallic components and detect weapons in security-related applications~\cite{8586961Li}.

Recently, the technological revolution of THz systems has given rise to highly autonomous machines and robots that merely require minor and occasional human intervention. However, the present industrial systems' high precision and efficient manufacturing processes necessitate advanced instantaneous control devices. The fast evolution of these automated systems requires ultra-fast data rates of terabit-per-second (Tbps). In light of this, deploying THz networks in practice can successfully meet such high data rate requirements. Moreover, the use of THz technology brings forth many potential sensing applications, such as holographic teleportation \cite{articleChaccour} and extended reality \cite{inproceedingsChaccour}, by delivering the extremely high reliability and low latency needed. According to the sixth generation (6G) of wireless networks' vision, leveraging THz systems to versatile networks is becoming a demanding feature on several emerging applications that require sensing, communications, and localization capabilities \cite{AKYILDIZ201416Akyildiz,sarieddeen2020overview}. Instead of solely relying on communication signals, such networks can use sensing signals to acquire situational awareness \cite{sarieddeen2019next, articleChaccour}, providing the estimated position, orientation, and state of target physical objects. Note that wireless sensing, in practical joint communications and sensing, concerns localization functionalities, such as radar sensing. However, the scope of this work reflects particularly on novel THz-specific conventional material (solids and gases) sensing.

\subsection{Comparison with infrared, microwave, and mmWave bands}

The THz-band's sensing capabilities are superior to their neighboring microwave and infrared counterparts. In particular, THz spectroscopy allows faster analysis with higher precision than infrared spectroscopy, which is further degraded by aerosol-induced light scattering and weather conditions \cite{articleHsieh}. Furthermore, many materials are transparent in the THz range. In contrast, most materials and living tissues are opaque in the microwave region, where the spectral data produces lower-definition images, despite being less sensitive to the weather conditions (due to larger wavelengths). Spectral data can reconstruct higher resolution images at arbitrary directions in the THz band due to antenna array beamforming capabilities and reduced scattering loss \cite{AKYILDIZ201416Akyildiz}. Compared to the fairly developed millimeter-wave (mmWave)-band technology \cite{9782674Chen}, the THz band offers better sensing capabilities and spatial resolution, but at the expense of shorter propagation distances. Furthermore, it is more convenient to characterize and spectroscopically analyze concealed contents, such as metallic and explosive materials and illicit drugs at the THz band, for which comparable spectral fingerprints do not exist in the mmWave region. This advantage is particularly important for applications requiring low vapor pressure sensing materials, such as explosive traces, that are hard to detect in the vapor-detection-based systems.

THz sensing and imaging also have shortcomings. For instance, blackbody radiation is dominant at THz frequencies at room temperature, complicating imaging-based object detection. THz signals are also severely absorbed by water vapor, resulting in gas identification being dominated by water vapor characteristics and material identification at long distances being very challenging under humidity. Moreover, the time response of a material in the reflected or absorbed THz radiation is calibrated by a sampling technique linked to the THz spectroscopy's pulsed nature, which limits its resolution. For temporal waveform acquisition, the spectral resolution depends on the temporal window inverse, and a trade-off between spectral resolution and temporal measurement time is thus unavoidable \cite{articleJepsen}. Such promising gains and shortcomings motivate research on signal processing techniques for THz sensing.

\subsection{Advancements in THz devices}

\begin{figure}[t]
\centering
\includegraphics[width=3.5in]{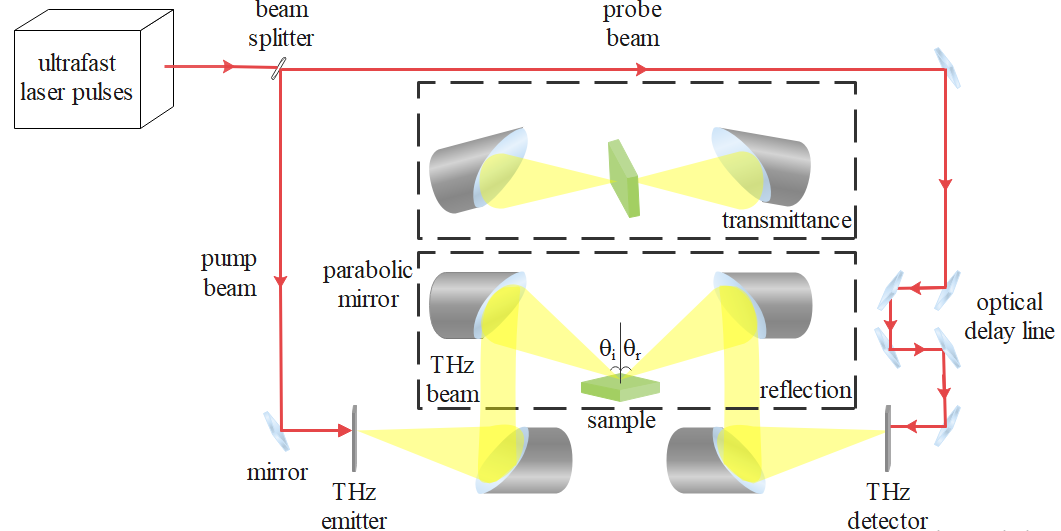}\vspace{-0.1in}
\caption{A conventional layout of a THz-TDS setup with reflection geometry and transmission geometry.}
\label{THzTDS}
\end{figure}

A typical THz system comprises a source, a detector, and intermediate optical components, with some systems consisting of two or more of these elements. The source produces broad-spectrum THz radiation. The components, i.e., lenses, mirrors, waveguides, and polarizers, manipulate the radiation. Then, a detector sensitive to THz rays measures the radiation reaching it as depicted in Fig. \ref{THzTDS}. THz systems are divided into two categories: Pulsed-wave and continuous-wave. Both systems are suitable for spectroscopic applications in imaging and sensing. Pulsed THz systems generate short broadband or near-single-cycle THz signals, whereas continuous-wave THz systems generate narrow-linewidth and frequency-tunable THz waves. Continuous-wave electronic sources include frequency-multiplied Gunn diode oscillators and backward wave oscillators (BWOs), whereas promising THz photonic sources include photomixing techniques and cryogenically cooled quantum cascade lasers (QCL) \cite{sengupta2018terahertz}.

THz-time domain spectroscopy (THz-TDS) is a powerful spectroscopic tool that has long been utilized to characterize and identify materials. Photo-conductive generation and detection of short, broadband THz pulses are at the heart of THz-TDS systems. An ultrafast laser emits femtosecond optical pulses, which are converted into picosecond THz pulses, and photoconductive antennas serve as the emitter and detector. The generation and detection techniques mainly depend on the application.

\subsection{Significance of signal processing for THz sensing}

The literature lacks a holistic approach for THz signal processing for communications and sensing, despite the progress made on THz system design. It is still unclear whether we can mitigate quasi-optical THz propagation to mimic favorable microwave propagation characteristics and enable seamless connectivity and sensing. Consequently, the questions of whether THz sensing can extend beyond application-specific settings and whether joint THz sensing and communications are of real practical value remain open at this early stage of research on the topic.

The progress on THz channel modeling paves the way for solid research in signal processing techniques for THz wireless sensing. Such research is particularly significant because several challenges have not yet been addressed to achieve THz technology's full potential. For example, THz signals are susceptible to high propagation losses and molecular absorption losses, which severely limit the achievable communication distances. Therefore, THz communications should be complemented by infrastructure-level and algorithmic-level enablers. At the infrastructure level, emerging beyond-5G technologies such as intelligent reflecting surfaces (IRSs) and ultramassive multiple-input multiple-output (UM-MIMO) configurations \cite{faisal2019ultra} are vital to expand the THz communication gains and overcome the distance problem. At the algorithmic level, novel signal processing techniques can optimize infrastructure utilization to get around THz quasi-optical propagation limitations and realize seamless wireless sensing. Efficient signal processing techniques are crucial to mitigate the noise and hardware impairments that distort measurements and worsen the sensing performance. This paper, therefore, presents a holistic overview of signal processing and machine learning techniques for efficient THz sensing, with an emphasis on signal pre-processing, feature extraction, and classification techniques. The paper further highlights the role of deep learning techniques by exploring their promising sensing capabilities at the THz band.

\subsection{Contributions of this work}

Several review papers on THz sensing exist in the literature, such as \cite{ruggiero2020invited,qin2013detection,yin2016application,ELHADDAD201398}. Up to the authors' knowledge, however, this paper is the first article that presents a holistic overview of signal processing and machine learning techniques for efficient THz sensing. More specifically, the paper introduces a roadmap for future THz sensing use cases, with a special focus on the role of signal processing. We promote the importance of both THz-TDS and frequency-domain spectroscopy in future reconfigurable THz systems. The frequency-domain spectroscopy approach, particularly, introduces significant flexibility in carriers' choice, thus converging on sensing information with minimum required measurements. We address the effect of THz channels on sensing performance. We perform a numerical merit investigation in which simulations validate our analyses. Such scope is significantly different from the treatment of THz sensing in literature, which mainly assumes controlled laboratory scenarios or specific use cases. We note that, at this early stage of research on the topic, the paper generates comprehensive numerical simulations using realistic data from existing databases \cite{gordon2017hitran2016,heilweil2011thz}, which would pave the way towards physical experimentations in future studies. The paper's main contributions can be summarized as follows:

\begin{enumerate}

\item We overview the studied signal processing and machine learning techniques in detail, assuming a unified THz system model, which facilitates the analyses of performance and complexity trade-offs. The presented overview particularly covers pre-processing (standard normal variate normalization, min-max normalization, and Savitzky-Golay filtering), feature extraction (principal component analysis, partial least squares, t-distributed stochastic neighbor embedding, and nonnegative matrix factorization), and classification techniques (support vector machines, k-nearest neighbor, discriminant analysis, and naive Bayes).
\item We motivate the importance of machine learning solutions compared to conventional signal processing. We explore the unique aspects and challenges of machine learning techniques that can be used in future THz systems towards enabling joint sensing and communication capabilities.
\item We motivate the importance of deep learning compared to conventional machine learning, especially in autonomous joint communications and sensing applications, where THz systems can generate a lot of data very fast.
\item We comment on the performance and complexity trade-offs of different techniques to come up with clear recommendations.
\item We motivate and formulate joint THz communication and sensing system models. \item We underline the key opportunities, challenges, and practical design considerations in deploying Terahertz communication and sensing systems. We further discuss the potential of novel machine learning techniques in leveraging the THz band for future joint sensing and communication technologies.

\end{enumerate}

\subsection{Organization of the paper}

The remainder of this paper is organized as follows. THz-TDS use cases and the corresponding spectroscopy system models are overviewed in Sec.\ref{sec:THzTDS}. Sec. \ref{sec:SPtech} reviews the different conventional signal processing and machine/deep learning methods that rely on raw recovered measurements for the pre-processing, feature extraction, and classification analysis of THz spectral data for material sensing. Sec.~\ref{sec:perf} highlights the relative performance and complexity trade-offs of the illustrated techniques based on numerical simulations. The possible application scenarios and technologies empowered by machine/deep learning are then discussed in Sec.\ref{sec:MLforTHz}. We also motivate prospect use cases of joint THz communication and sensing systems and investigate their system design considerations in Sec. \ref{sec:JCAS}, detailing the corresponding challenges and opportunities and highlighting the importance of carrier-based spectroscopy. In Sec. \ref{future:res}, we discuss the open research directions that are envisaged to be at the forefront of THz sensing and communications technologies. We conclude the paper in Sec.\ref{sec:conc}. 

Regarding notation, bold upper case, bold lower case, and lower case letters correspond to matrices, vectors, and scalars, respectively. $\Prb(\alpha)$ is the probability that event $\alpha$ occurs. $(\cdot)^{T}$ and $(\cdot)^{-1}$ stand for the transpose and inverse, respectively. Scalar norms (or absolute values) are denoted by $\abs{\cdot}$ and Frobenius norms are denoted by $\norm{\cdot}$. The notation $j\!=\!\sqrt{-1}$ is the imaginary number.

\section{Terahertz-time domain spectroscopy}
\label{sec:THzTDS}

Spectroscopy is a reliable technique for material and gas identification over a wide range of spectral data, allowing the extraction of material parameters without requiring multiple samples of different thicknesses. THz-TDS \cite{articleJepsen} characterizes the refractive indices and absorption coefficients, which correspond to the static and transient optical properties of materials. THz pulses' temporal profiles are recorded twice with and without the media under examination in a conventional THz-TDS measurement. These pulses are referred to as the incident and transmitted pulses, or reference and sample pulses, respectively. The recorded pulse traces correspond to the electric field's temporal dependence associated with THz waves, the transient THz electric field at the detector. The two profiles are then Fourier-transformed to obtain their complex-valued spectral behavior, where the ratio of the two pulses' electric field strengths in the frequency domain determines the optical properties of the sample material. The complex spectrum yields amplitude and phase information and allows direct calculation of the frequency-dependent index of refraction, absorption coefficient, and sample thickness. Compared to conventional Fourier-transform-based infrared and Raman spectroscopy, THz-TDS infers more useful information and gives direct access to the electrical field amplitude.

\subsection{Transmission- and reflection-based THz-TDS}

The necessary spectroscopic measurements that provide the diagnostic tools for determining the optical constants of materials are transmission-based and reflection-based. Transmission spectroscopy is correlated with absorption spectroscopy, which is based on analyzing the amount of light absorbed by a sample material at a given wavelength. On the other hand, reflection spectroscopy studies the reflected or scattered light from a material as a function of wavelength. For both reflection and transmission measurements, the Fresnel equations describe the energy transfer at the interface between two media, usually air and another homogeneous material. Most THz sensing and imaging applications are conducted in a transmission geometry (absorbance THz spectroscopy), owing to its simple set-up design and the high contrast of the transmitted THz signal. 

Several reasons, however, make THz reflection-based TDS more appealing for material identification. Reflection geometry is capable of measuring the spectrum of thick or highly absorptive materials that are opaque in the THz band. In particular, reflection geometry makes detecting targets on non-transparent substrates possible, such as in the case of threat items concealed in opaque envelopes and paint on the bodies of vehicles. Furthermore, reflection spectroscopy is capable of reconstructing full three-dimensional (3D) images of objects, where the short THz pulse duration provides precise means of calculating the distance to reflecting surfaces. Despite providing low-contrast spectroscopic imaging, reflection geometry offers a unique solution to applications at long distances, such as standoff distance THz sensing and imaging in open field environments. The reflection configuration is also attractive for applications that require contact-free, nondestructive inspection, such as inspection of the coating thickness of materials and diagnostic studies of normal and cancerous biological tissues. Due to the broadband nature of THz signals and complex environments (cloud, rain, dust), inevitable scattering losses and attenuation degrade the collected THz spectral response and limit the transmission distance; reflection losses are also critical. However, several optical techniques exist to optimize both transmission and reflection geometries to enable accurate acquisition of the optical constants of materials. Such techniques include ultra-thin metal films, off-axis parabolic mirrors, and wire-grid polarizers. Leveraging a large number of antennas to introduce high beamforming gains can also mitigate propagation losses \cite{sarieddeen2020overview}.

\subsection{Gas, liquid, and solid THz-TDS}

THz-TDS supports inspection and identification of different states of matter through electromagnetic interactions with gas, liquid, and solid components:

\begin{enumerate}

\item Gas Spectroscopy \cite{articleHsieh}: Used to detect air pollution, flammable gases, atmospheric pressure, and humidity, or to monitor molecules blended in unfavorable aerosols like smoke, fog, haze, dust, fume, and others.

\item Liquid Spectroscopy \cite{George2013}: Used to characterize liquid and aqueous solutions. The dielectric properties of liquids depend on interactions and dipole relaxations. T-rays are sensitive to relaxational, oscillatory, and collective motions and can be used to study ionic liquid solutions. The high sensitivity to water and non-destructive penetration abilities of THz radiation are appealing for a plethora of liquid spectroscopy applications, such as the identification of water content, alcohol, liquid fuels, and petrochemicals. THz-TDS can also investigate the color, carbonation, and flavor of commercial beverages.

\item Solid Spectroscopy \cite{hangyo2005terahertz}: Used to investigate a wide variety of spectral fingerprints of molecular solids such as explosives, dielectrics, polymers, ferroelectrics, semiconductors, and photonic crystals, for security screening, quality inspection, or pharmaceutical applications. Several THz-TDS methods can quantify the concentration, index of refraction, absorption peak, dielectric constants, and polarizability of materials. There are also different kinds of low-energy excitations in solid materials that THz-TDS can successfully detect.

\end{enumerate}

\subsection{THz-TDS imaging}

In THz material sensing, the inspection is not consistently feasible if the samples under investigation have no apparent features such as absorption peaks. Nevertheless, the capability to form images and provide spectral information by exploiting other T-ray unique aspects enables alternative THz-based material characterization techniques. In particular, there has been a resurged interest in THz time-domain spectroscopic imaging because of the low power and low energy interaction of THz waves with some materials \cite{articleJepsen}. The image data and spectral information can determine the type and the shape of the material. For instance, a non-contact method uses THz pulse imaging to quantify the coated pharmaceutical tablets' thickness. By setting a THz-TDS system with an intermediate focus, one can measure the THz waveform of the pulse traversing an object placed at the focal plane. Because of low THz scattering, THz-TDS imaging systems produce high contrast images that enable effective material analysis.

There are several THz-TDS imaging methods, each of which can extract different relevant information from samples. Such methods are amplitude-based, phase-based, or a combination of the two. With amplitude imaging, the THz waveform's magnitude is measured using fast numerical Fourier transform over a particular frequency band to extract the waveform's peak signal at each pixel. To form an image pixel by pixel, the transmitted THz wave is measured for each position of the object under study. After THz signal acquisition, THz imaging requires the processing of the entire waveform. Each waveform holds a wealth of information (both in amplitude and phase) that can be extracted at each image pixel with integrated digital signal processors; detailed chemical or physical information can be obtained at each pixel. In our paper's context, signal processing for THz sensing does not account for imaging techniques, which we rather consider an extension of this current work.

\subsection{Reflection spectroscopy system model}

THz reflection spectroscopy can measure the real and imaginary parts of both the refractive index and dielectric constant of sample material to obtain distinct absorption coefficients. The Kramers-Kr\"{o}nig transform technique \cite{HERRMANN2012107} can be applied to calculate the sample's optical conductivity, and the reflection coefficient can be determined using Fresnel's equation. When a THz beam propagating in a uniform medium (air) encounters a material with refractive index $n_i$ at incident angle $\theta_i$, the ratio of the reflected beam $E_r$ to the incident beam $E_0$ yields the complex reflectivity of the s-polarized component (denoted $\tilde r_s$) and the p-polarized component (denoted $ \tilde r_p$). We have
\begin{equation}
   \tilde{r}_{s}=\frac{{E}_{rs}}{{E
  }_{0s}}= \frac{\tilde{n}_i \cos\theta_{i} -n_r \cos\theta_{r}}{n_r \cos\theta_{r} +\tilde{n}_i \cos\theta_{i}} ,
  \end{equation}
  \begin{equation}
   \tilde{r}_p=\frac{{E}_{rp}}{{E}_{0p}}= \frac{\tilde{n}_i \cos\theta_{r} -n_r \cos\theta_{i}}{n_r \cos\theta_{i} +\tilde{n}_i \cos\theta_{r}},
\end{equation}
where $E_{rs}$ and $E_{0s}$ are the reflected and incident beams of the s-polarized component, and $E_{rp}$ and $E_{0p}$ are the reflected and incident beams of the p-polarized component, respectively. The refractive index of air is $n_r=1$, and for both polarizations we have $n_i \sin\theta_i\!=\!n_r \sin\theta_r$ (the incident and reflected angles are related via Snell's Law). Reflection at normal incidence facilitates deriving the reflection coefficient through the complex refractive index of the material $\tilde{n}\!=\!n\!+\!j \mathcal{X}$, with ${ \mathcal{X}}$ being the extinction coefficient, and where
\begin{equation}
   { \tilde{r}}= \frac{{E}_r}{{E}_0}=\frac{(n-1)+j \mathcal{X}}{(n+1)+j \mathcal{X}}=\sqrt{{R}}e^{j\phi}.
\end{equation}
Therefore, the reflection coefficient is $r\!=\!\sqrt{{R}}$, where $R$ is the reflectance. Assuming both the phase shift $\phi$ and ${R}$  are known, the refractive index and the extinction coefficient are calculated as
\begin{equation}
n=\frac{1+{R}}{1+r-2\sqrt{{R}}\cos\phi},
\end{equation}
\begin{equation}
    \mathcal{X}=\frac{2\sqrt{{R}}\sin\phi}{1+{R}-2\sqrt{{R}}\cos\phi}.
\end{equation}
The absorption coefficient can then be computed as
\begin{equation}
    \mathcal{K}=\frac{4\pi f \mathcal{X}}{c},
\end{equation}
where $f$ is the frequency, and $c$ is the speed of light in vacuum.

\subsection{Transmission spectroscopy system model}

Transmission THz-TDS probes collective vibrational modes in materials. In particular, the frequency-dependent complex refractive index can be measured entirely from the time-dependent electric field of the THz waveform without the need for Kramers-Kr\"{o}nig transform equations \cite{HERRMANN2012107}. Using the sample thickness $d$ and the complex refractive index $\tilde{n}$, the amplitude of THz transmittance, denoted by $T$, can be obtained as

\begin{equation}
    {T}= \frac{{E}_t}{{E}_0}\approx\frac{4n}{(1+n)^2}e^{\frac{j2\pi f(\tilde{n}-1)d}{c}},
\end{equation}
where $E_0$ and $E_t$ are the incident and transmitted amplitudes, respectively. Consequently, we can obtain the extinction coefficient $\mathcal{X}$, and refractive index $n$ as
\begin{equation}
    \mathcal{X}= \frac{c}{2\pi df}\ln\left(\frac{4n}{{T}(1+n)^2}\right),
\end{equation}
\begin{equation}
     n=\frac{c\phi}{2\pi d f}+1,
\end{equation}
where $\phi$ is the phase difference between the reference
and sample. The absorption coefficient of the sample material can then be calculated as
\begin{equation}
  \mathcal{K}=\frac{2}{d}\ln\left(\frac{4n}{{T}(1+n)^2}\right).
\end{equation}

In gas spectroscopy, the absorption coefficient can be extracted from the path gain, which is expressed as
\begin{equation}\label{eq:LoS}
	\alpha =   \frac{c}{4\pi f D} \times e^{ -\frac{1}{2} \mathcal{K}(f) D }  e^{  -j \frac{2\pi f}{c} D},
\end{equation}
where $D$ is the distance between the transmitter and receiver.
$\mathcal{K}(f)$ can be expressed as the sum of absorption contributions from isotopes ($i\!\in\! {1,\cdots,I}$) of gases ($g\!\in\! {1,\cdots,G}$) that the medium is composed of, and it is expressed as
\begin{equation}
    \mathcal{K}(f)=\sum_{i,g}\mathcal{K}^{i,g}(f).
\end{equation}
Using radiative transfer theory \cite{Jornet5995306}, the molecular absorption coefficient for an individual isotope can be expressed as
\begin{equation}
    \mathcal{K}^{i,g}(f)=\frac{p}{p_{0}}\!\frac{t_{\mathrm{STP}}}{t}\! \frac{p}{g_ct} q^{i,g\!} \,A_c \sigma^{i,g},
\end{equation}
where $p$ is the system pressure, $p_{0}$ is the reference pressure, $t$ is the temperature, $t_{\mathrm{STP}}$ is the temperature at standard pressure, $g_c$ is the gas constant, $q^{i,g}$ is the mixing ratio for the isotope of gas ($i,g$), $A_c$ is the Avogadro constant, and $\sigma^{i,g}$ is the absorption cross section that is given as
\begin{equation}
   \sigma^{i,g}= F^{i,g\!}(f)\,S^{i,g\!}\, \frac{f}{f_{c}^{i,g\!}} \frac{\text{tanh}\left(\frac{P_ccf}{2K_Bt}\right)}{\text{tanh}\!\left(\frac{P_cc\left(f_{c}^{i,g\!}\right)}{2K_Bt}\right)},
\end{equation}
where $F^{i,g\!}(f)$ is the Van Vleck-Weisskopf line shape \cite{Huang17}, $S^{i,g\!}$ is the absorption strength, $P_c$ is the Planck constant, and $K_B$ is the Boltzamnn constant. $f_{c}^{i,g\!}\!=\!f_{c0}^{i,g\!}\!+\!\delta^{i,g\!}\frac{p}{p_0}$ is the resonant frequency of the isotope of gas $(i,g)$, where $f_{c0}^{i,g\!}$ is the zero-pressure of the resonant frequency and $\delta^{i,g\!}$ is the linear pressure shift. The Van Vleck-Weisskopf line can be further decomposed as a function of Lorentz half-width $w_L^{i,g}$ for gas $(i,g)$ as follows:
\begin{equation}
    F^{i,g\!}(f)\!=\!\frac{cw_L^{i,g}{f}}{\pi{f_{c}^{i,g\!}}}\left[\frac{100}{(f\!-\!f_{c}^{i,g\!})^2\!+\!(w_L^{i,g})^2}\!+\!\frac{100}{(f\!+\!f_{c}^{i,g\!})^2\!+\!(w_L^{i,g})^2}\right].
\end{equation}
The Lorentz half-width ($ w_0^{i,g}$) \cite{Jornet5995306} can be obtained as a function of the broadening coefficient of both air ($w_0^{\mathrm{air}}$) and the isotope $i$ of gas $g$ as
 \begin{equation}\label{airAtt}
     w_L^{i,g}=\left[\left(1\!-\!q^{i,g\!}\right)w_0^{\mathrm{air}}\!+\!q^{i,g\!}w_0^{i,g\!}\right]\!\left(\frac{p}{p_0}\right)\!\left(\frac{t_0}{t}\right)^\gamma,
 \end{equation}
where $\gamma$ is the temperature broadening coefficient parameter.
Note that all the aforementioned parameters can be extracted from the high-resolution transmission molecular absorption database (HITRAN) database \cite{gordon2017hitran2016}. Although the complexity of the radiative transfer theory approach is high, the existing alternative gas absorption models are not accurate enough for sensing purposes; they are rather approximations that can be used in THz communication scenarios \cite{sarieddeen2020overview}.

In our system model, for material sensing, we assume the THz transmission spectra to be accumulated in a vector $\mbf{x}^{\mathrm{Tr}}\!=\![{x}_{1}^{\mathrm{Tr}} {x}_{2}^{\mathrm{Tr}}\cdots {x}_{N}^{\mathrm{Tr}}]\!\in\!\mathcal{R}^{N\times1}$, where $x_i^{\mathrm{Tr}}\!=\!T(f_i)$ represents the input data. For the gas sensing scenario, the absorption coefficient spectra is accumulated in a vector $\mbf{x}^{\mathrm{Abs}}\!=\![{x}_{1}^{\mathrm{Abs}} {x}_{2}^{\mathrm{Abs}}\cdots {x}_{N}^{\mathrm{Abs}}]\!\in\!\mathcal{R}^{N\times1}$, where $x_i^{\mathrm{Abs}} \!=\!\mathcal{K}(f_i)$ is used as input data. In the remainder of this paper, we drop the superscripts $\mathrm{Abs}$ and $\mathrm{Tr}$ when the discussion equally applies for both absorption and transmission spectra, for simplicity. We next detail the signal processing and machine learning techniques that can be applied on $\mbf{x}$ to retrieve sensing information. We first start with overviewing the pre-processing techniques in Sec. III. Then, we present the feature extraction techniques, quantitative analysis of materials, and deep learning prospects in Sec. IV, Sec. V, and Sec. VI, respectively. Table \ref{table:log} lists sample works from the literature that utilize a variety of signal processing techniques for different THz sensing applications.

\section{Conventional Machine Learning Candidates for THz Sensing }
\label{sec:SPtech}
\subsection{Signal pre-processing techniques}

Perfect recovery of the THz signal at the detector is vital in spectroscopic systems. However, several factors prevent such ideal conditions. For instance, atmospheric attenuation significantly influences the propagating THz radiation in the air. Consequently, loss of information, slope variations, baseline shifts, and redundancy in the acquired data are expected. The quality of classification depends not only on the effectiveness of the feature extraction techniques but also on the pre-processed data's quality and quantity. Employing signal pre-processing techniques for spectroscopic THz systems can thus significantly improve the accuracy, noise robustness, and resolution of THz sensing and imaging systems.

We denote the spectral output after applying pre-processing techniques on an input $\mbf{x}$ by the vector $\hat{\mbf{x}}\!=\![\hat{x}_{1} \hat{x}_{2}\cdots \hat{x}_{N}^{}]\!\in\!\mathcal{R}^{N\times1}$. Out of a vast range of signal pre-processing techniques, a few methods have been more commonly applied to THz spectral data, namely, Savitzky-Golay (SG) filtering, de-trending (DT), first derivative (FD), standard normal variate (SNV), baseline correction (BC), wavelet transform, and min-max normalization. Other spectral pre-treatment techniques have also been studied to pre-process THz and near-infrared (NIR) spectra, such as the logarithmic function $\log(1/R)$, mean centering (MC), and multiplicative scatter correction (MSC). In this work, we detail and implement SNV, min-max normalization, and SG filtering.

\subsection{Feature extraction techniques}

Inconclusive classification results can be obtained in THz spectroscopy due to unquantifiable scattering effects which form spurious structures. Towards mitigating such THz-TDS problems, following signal pre-processing, we target extracting critical features from the THz spectral data. In particular, we consider the following feature extraction techniques: Principal component analysis (PCA), t-distributed stochastic neighbor embedding (t-SNE), non-negative matrix factorization (NMF). Following our system model, feature extraction techniques operate on an initial set of THz spectral data in a matrix $\mbf{X} \!\in\! \mathcal{R}^{M\times N}\!=\! [\mbf{x}_1, \mbf{x}_2, \cdots, \mbf{x}_{M}]^T$, or a pre-processed matrix $\hat{\mbf{X}}$, where $M$ denotes the number of data samples (observations) and $N$  corresponds to a range of THz features (variables) that are subsequently used as inputs to the classifiers. The dataset matrix is expressed as
\begin{equation}
\mbf{X}=
\begin{pmatrix}
x_{1,1} & x_{1,2} & \cdots & x_{1,N} \\
x_{2,1} & x_{2,2} & \cdots & x_{2,N} \\
\vdots  & \vdots  & \ddots & \vdots  \\
x_{M,1} & x_{M,2} & \cdots & x_{M,N}
\end{pmatrix},
\end{equation}
and the ouptut feature vector is expressed as $\mbf{y}\!=\![y_1, y_2, ..., y_P]$, where $P\!\leq\! N$.
\begin{enumerate}
\item{Principal Component Analysis: }
PCA is a dimensionality reduction technique that reduces a multi-dimensional dataset of many correlated variables into a smaller set with few comprehensive indicators. The new indicators are known as principal components (PCs). In THz material classification, PCA is abundantly used as an unsupervised, non-parametric method for extracting relevant data features and eliminating overlapping data from original THz spectral datasets. The matrix $\mbf{X}$ is converted into a vector of synthesis indicators (PCs), $\mbf{y^{\mathrm{PCA}}}\!=\![y_1^{\mathrm{PCA}}, y_2^{\mathrm{PCA}}, ..., y_P^{\mathrm{PCA}}]$, the entries of which are sorted in descending order of their respective variances. The PCs correspond to the eigenvectors of the $P$ larger eigenvalues of the covariance matrix $\mbf{C}\!=\!\mbf{X}^T\mbf{X}/P-1$.

\item{t-Distributed Stochastic Neighbor Embedding: }
t-SNE is a non-linear learning method that reflects low-dimensional data by optimally positioning data points in a projection map. The t-SNE algorithm utilizes the joint probability distribution between high-dimensional data points and their corresponding synthetic data points in a low-dimensional space, minimizing the Kullback-Leibler (KL) divergence to obtain optimal low-dimensional data.

For the same input $N$-dimensional data $\mbf{x}$, the similarity conditional probabilities are first computed as
\begin{equation}
\Prb(j|i)=\frac{\mathrm{exp}\left(-\abs{x_j-x_i}^2/2{\sigma_i}^2\right)}{\sum_{k\neq j}\mathrm{exp}\left(-\abs{x_j-x_k}^2/2{\sigma_i}^2\right)}
\end{equation}
where $\Prb(j|i)$ is the similarity of data point $x_j$ to data point $x_i$ in $\mbf{x}$, and $\sigma_i$ is the variance of the Gaussian-distributed THz spectral value centered over $x_i$. Using a heavy-tailed Student t-Distribution with one degree of freedom in the low-dimensional space, the joint probability $\mathrm{Qr}(j|i)$ between synthetic data points in $\mbf{y}$ is calculated as
\begin{equation}
    \mathrm{Qr}(j|i)=\frac{\left(1+\abs{y_i-y_j}^2\right)^{-1}}{\sum_{k\neq i}\left(1+\abs{y_i-y_k}^2\right)^{-1}}.
\end{equation}

Then, the $\mathrm{KL}$ divergence between the synthetic data points in low-dimensional space and their corresponding data points in high-dimensional space is minimized as
\begin{equation}
\mathrm{Ct} = \mathrm{min} \sum_i\sum_j \Prb(j|i)\log\frac{\Prb(j|i)}{\mathrm{Qr}(j|i)},
\end{equation}
with $\Prb(i|i)\!=\!\mathrm{Qr}(i|i)\!=\!0$.

\begin{figure*}[t]
\centering
\includegraphics[width=6.8in]{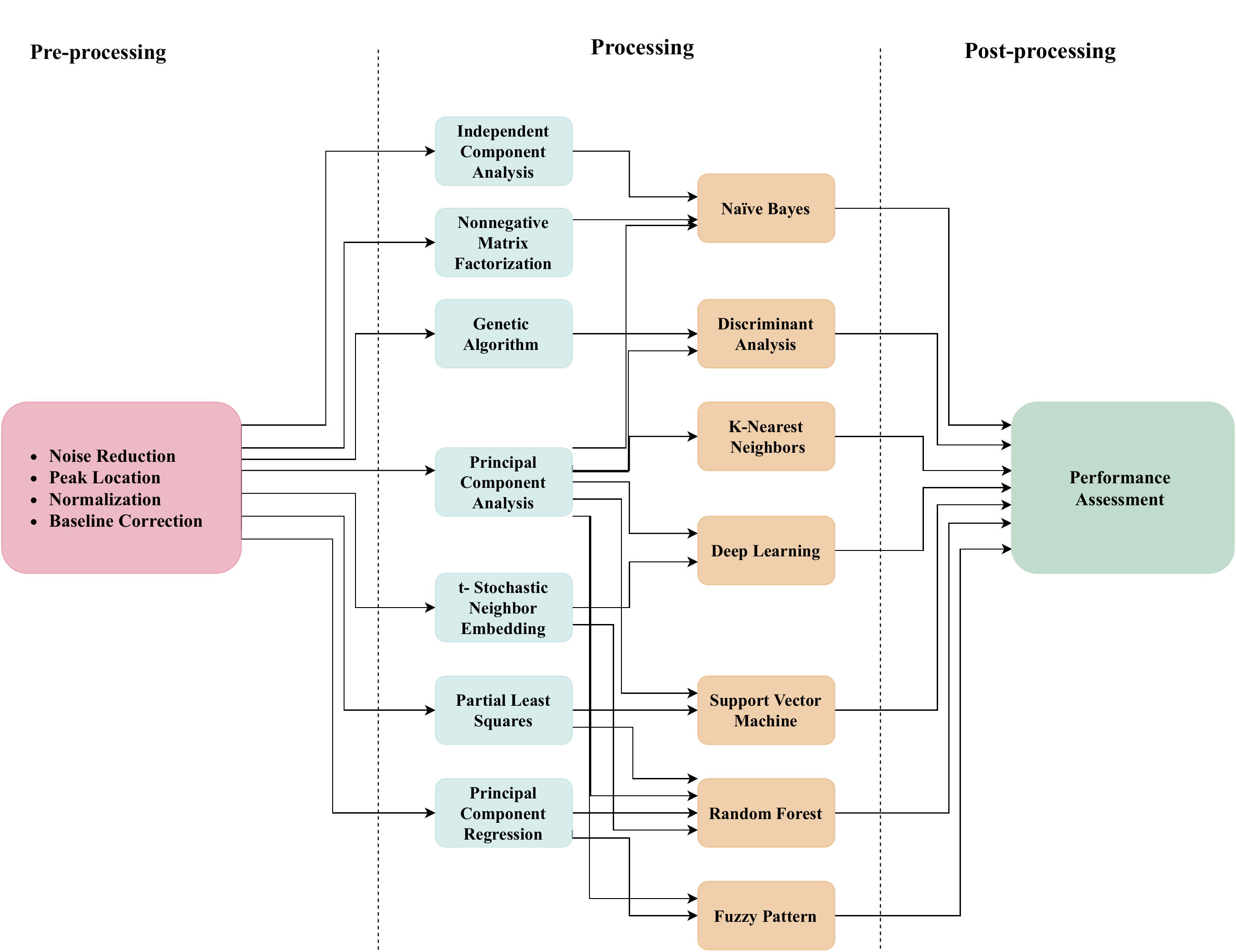}

\caption{A map of candidate machine learning techniques for THz sensing.}
\label{f:table}
\end{figure*}

\begin{table*}[htbp]
\centering
\caption{THz-TDS Chemometrics }\label{table:log}

\begin{tabular}{ |p{3.5cm}|p{1.8cm}|p{4.5cm}|p{1.8cm}|p{2.2cm}|p{1cm}| }
 \hline
       Materials & THz Range & THz-TDS Setup Details& Feature Extraction & Classification & Ref. \\
       \hline
       \hline
          Aminoacids, saccharides, and inorganic substances & 0.9-6 THz & GnAs photoconductive antenna, femtosecond laser, transmission mode&PCA & Fuzzy Pattern & \cite{10.1117/12.527880}  \\
       \hline
        Flavonols (myricetin, quercetin, and kaempferol)& 0.6-2.7 THz &{Mode-locked Ti-sapphire laser, transmission mode} &PLS& Random forest, LS-SVM &\cite{articleYan}\\
       \hline
         Rice&0-6.4 THz &TERA K15 All fiber-coupled spectrometer, femtosecond laser, transmission mode& PCA& PLS-DA, SVM, BPNN& \cite{8977502Li } \\
       \hline
         Protein (bovine serum albumin) & 0.2-1.2 THz&LT-GaAs photoconductive antenna, femtosecond laser, lock-in amplifier & PCA, t-SNE & Random forest, NB, SVM, XGBoost & \cite{articleCan}\\
       \hline
        Benzoic acid &1.6-2.8 THz &TAS7500SU system, transmission mode, ultra-short pulse fibre lasers &PCA&GRNN, BPNN &  \cite{190485Xudong}\\

 \hline
  Pure analytes (citric acid, fructose, and lactose)&0.05-3 THz & TPS Spectra 3000, mode locked Ti Sapphire laser, transmission mode &PCA&PLSR, ANN & \cite{026004Bowman} \\

 \hline
 Transgenic rice and Cry1Ab protein&0.1-2.6 THz & Z-3 THz-TDS system, LT GaAs photoconductive antenna, ZnTe electro-optical crystal detector &PCA&PLSR, DA &  \cite{articleWendao} \\

 \hline
  Rice and imidacloprid pesticide  &0.3-1.7 THz & mode-locked Ti-sapphire laser, femtosecond laser,
 ZnTe photoconductive antenna,
 transmission mode&PLS&SVR  & \cite{CHEN20151}\\

 \hline
Aflatoxins B1 in acetonitrile solution   &0.4-1.6 THz & mode-locked Ti:sapphire laser, photoconductive switches,
 transmission mode& PCA, PLS, PCR&SVM  &  \cite{GE2016286}\\

 \hline
  Fuel oils (lubricant, gasoline, and diesel) & 0.2-1.5 THz& mode-locked femtosecond Ti-sapphire laser, lock-in amplifier, transmission mode &PCA&SVM, BPNN & \cite{Zhan_2016} \\

 \hline
 Extra-virgin olive oil (EVOO)  &0.1-4 THz &   TAS7500TS HF THz-TDS system, femtosecond laser, transmission mode &PCA, Genetic algorithm &LS-SVM, BPNN, Random Forest &  \cite{LIU201886} \\

 \hline
 Adulterated dairy products (skim, low fat milk)&0.1-1.5 THz & \centering\textemdash &PCA&SVM-DA  &  \cite{articleLiu}\\

 \hline
 Oral lichen planus (OLP)&0.3-3.5 THz &T-SPEC THz spectrometer, absorption mode  &PCA& SVM&  \cite{045001Yury}\\

 \hline
\end{tabular}
\centering
\end{table*}

\item{Non-negative Matrix Factorization: }
NMF is another dimensionality reduction and feature extraction technique suitable for high-dimensional multivariate THz spectral data analyses. NMF approximates the data by iterative additive combinations of the basis vectors, making it a good candidate when other tools can not guarantee non-negativity in measurements that contradict physical realities. For a non-negative matrix $\mbf{X} \in \mathcal{R}^{M\times N} $ and a positive integer factorization rank $P \!\ll\! \mathrm{min}(N,M)$, we find two non-negative matrices, $\mbf{W}\! \in \!\mathcal{R}^{M\times P}$ and $\mbf{H} \!\in \!\mathcal{R}^{P\times N}$, the product of which approximates $\mbf{H}$ via non-negative factorization:
\begin{equation}
    \mathrm{min}\norm{\mbf{X}-\mbf{W}\mbf{H}}^{2}.
\end{equation}
The number of columns in $\mbf{W}$ is the latent feature representing the reduced feature space dimension.
\end{enumerate}

\subsection{Classification techniques}

Following the discussion on feature extraction, we investigate complementing machine learning techniques that classify materials based on their THz spectral absorption and transmission coefficients. The candidate combinations of signal processing techniques are illustrated in Fig. \ref{f:table}. We next detail the following candidate classification techniques: Naive Bayes (NB), support vector machine (SVM), K-nearest neighbor (KNN), and partial least squares-discriminant analysis (PLS-DA). Unless otherwise stated, for an input set $\mbf{y}\!=\! [y_{1}, y_{2}, \cdots, y_{P}]$ of $P$ features, the vector $\mbf{w}\!=\! [w_{1}, \cdots,w_{m}, \cdots, w_{k}]$ denotes the output data classes.

\begin{enumerate}
\item{\textit{Naive Bayes: }}

The NB classifier is a supervised probabilistic machine learning classifier that applies the maximum a posteriori (MAP) decision rule for parameter estimation. NB assumes the presence and absence of each feature of a class independently. Each class has a probability $\Prb({w}_{m})$ that is estimated from the training feature dataset. The class with the highest post-probability is the resulting target class. Using Bayes' theorem, the NB classifier computes the conditional probability $\Prb({w}_{m}|\mbf{y})$ for each of the $K$ possible classes as
\begin{equation}
     \Prb({w}_{m}|\mbf{y})=\frac{1}{\Prb(\mbf{y})}\Prb(w_m)\prod_{i=1}^{p}\Prb(y_i|w_m).
\end{equation}
 For the final decision, the classifier model incorporates a MAP rule to predict the class with the largest posterior probability:
\begin{equation}\label{NBeqn}
    w_m=  \underset{w_m}{\mathrm{argmax}} \Prb(w_m)\prod_{i=1}^{p}\Prb(y_i|w_m).
\end{equation}

The NB classifier can be used alongside 2-D cross-correlation for classifying THz signals. The correlation between the background time-domain pulse and the sample ensures considerable noise suppression in a THz dataset. Consequently, any phase differences due to sample dispersion are preserved and discriminated between the two signals.

\item{\textit{Support Vector Machine:}}

SVM is a linear regression model that classifies data based on a set of support vectors, subsets of the training dataset, that construct a hyperplane in feature space. For both linear and nonlinear problems, SVM classifies data using a boundary hyperplane that separates data into different classes. Most material recognition problems consisting of multi-class pulsed signals (signals belonging to three or more classes) use SVM classifiers to analyze and discriminate data.

A set of learning data $(y_1,w_1), \cdots, (y_l,w_l),\cdots, (y_L,w_L)$ is used for building the SVM model, where $\mbf{y}_l\in \mathcal{R}^N$ and $w_l \in (1, 2, ..., k)$  denotes the class label corresponding to each input feature vector $\mbf{y}_l$. A total of $\frac{k(k-1)}{2}$ classifiers are constructed, where $k$ denotes the class number of the input data. Each classifier is trained on input data from two classes. By training data from the $\nth{i}$ and the $\nth{j}$ classes, the classification problem can be expressed as
\begin{equation}
   \underset{\mbf{a}_{ij},b_{ij},\xi_{ij}}{\mathrm{min}}\frac{1}{2}\mbf{a}_{ij}^T\mbf{a}_{ij}+c\sum_{t}\xi_t,
\end{equation}
Subject to
\begin{equation*}
\begin{aligned}
    \mbf{a}_{ij}^T\phi(y_t)+b_{ij}\geq1-\xi_{ij}, \mathrm{if} \: w_t=i,\\
    \mbf{a}_{ij}^T\phi(y_t)+b_{ij}\leq-1-\xi_{ij}, \mathrm{if} \: w_t=j,\\
    \xi_{ij}\geq0.
    \end{aligned}
\end{equation*}
where $\mbf{a}$ is the normal vector of the hyperplane, $b$ is a real-valued bias, $\xi$ is the slack variable, $c$ is the penalty parameter, $t$ is the index of the combined set of the $\nth{i}$ and $\nth{j}$ samples of the training data, and $\phi(.)$ is the function that maps the training data (input space) to a higher dimensional space (feature space). Both $\mathbf{a}$ and $b$ denote the optimization variables for the optimal hyperplane.

\item{\textit{K-Nearest Neighbor (KNN): }}

KNN is a distance-based learning algorithm that is favorable due to its simple mathematical formulation and relatively little training time.  In KNN, data points close to each other are referred to as neighbors, and the desired class is constructed based on distances to the data points near known data. The algorithm acquires $K$ nearest neighbors by a majority vote decision, based on a specific distance metric (Euclidean, Mahalanobis, Chebychev, or correlation distance). The controlling variable $K$ is chosen after preliminary validation or hyperparameter optimization, depending on the dataset requirements. The KNN algorithm selects the predicted class for which the distance to the test data is minimized.

\item{\textit{Partial Least Squares-Discriminant Analysis: }}

PLS-DA is derived from the PLS regression (PLS-R) algorithm and combines the properties of PLS-R with the discrimination power of a classification technique. PLS-DA thus sharpens and maximizes the separation between groups or classes of observations. PLS is a commonly used supervised feature extraction technique for THz data. It reduces data in a low dimensional space via linear transformation. However, it forms a hybrid classification method when combined with DA, which can be used for predictive modeling (we thus discuss PLS-DA under classification techniques). The fundamental PLS-DA paradigm is discussed in \cite{8977502Li}.

\end{enumerate}

\subsection{Deep learning}

THz spectral data sets can be extensive and complex, so neural networks can be explored as robust classification tools to speed up learning efficiency compared to conventional classification models discussed earlier in this paper. We distinguish various neural network architectures depending on the approach of network training and classifying. In particular, deep learning neural network techniques can be supervised, unsupervised, and reinforced. One of the main advantages of neural networks is that they can create new features by themselves, unlike traditional shallow learning techniques in which features need to be identified accurately by other techniques. Therefore, deep learning classifiers can operate directly on THz training data, enabling faster learning that is much needed in fast-changing THz conditions of contemporary and future wireless communication networks. Such conditions make sensing paradigms slowly responsive to environmental changes, thus resulting in degraded performance. In static sensing at THz frequencies, e.g., close-distance sensing, the fast-changing conditions can be due to changing molecular components in the cross-section of narrow beams or, more likely, due to the narrow beam itself, where a simple beam sweep can change channel characteristics. In dynamic sensing, e.g., Vehicle to Everything (V2X)-aided sensing, the fast-changing conditions can be due to the multidimensional communications medium and evolving regulatory environment. In this context, deep learning can be a potential enabling solution to the parameter adaptation issues of traditional learning methods. 

Supervised and unsupervised deep learning techniques can be further realized in JCAS scenarios, providing the user's instantaneous location, orientation, and velocity information (this can be achieved using the high-resolution estimated THz sensing parameters in conjunction with the network communication data). Furthermore, in terms of training and testing accuracy, deep learning outperforms simpler machine learning algorithms in THz applications that deal with high dimensional spatio-temporal data and traffic prediction/classification using clustering algorithms. Along this line, advances in machine learning techniques have also introduced novel learning models, such as deep reinforcement and transfer learning, providing even more enhanced performance and training speed. Although deep learning models achieve excellent accuracy and efficiency, they require large datasets of labeled samples. To counterbalance this problem, transfer learning is often adopted, in which pre-trained models can be used to speed up the training and learning efficiency. This is particularly valuable in THz-based deep learning applications, as most neural networks exhibit significant training latency using large and challenging THz datasets. 

We next highlight two particular types of supervised neural networks because of their popular usage in existing THz material sensing literature: Generalized regression neural networks (GRNN) \cite{sun2019generalized} and backpropagation neural networks (BPNN) \cite{ye2020characterization}.
\begin{enumerate}
\item{\textit{Generalized Regression Neural Network: }}

The most extensively used deep learning model in THz material classification studies is GRNN, both for theoretical and practical applications \cite{sun2019generalized}. GRNN is a typical feed-forward neural network that provides a powerful variation to the conventional radial bases function neural network. As a single-pass efficient learning technique, GRNN solves tedious efficiency and flexibility issues. Unlike BPNN, GRNN requires selecting only one training parameter to be learned (the smoothing factor or the width of radial basis functions, for example). The corresponding performance differs significantly with smoothing factor choice, where the estimated density takes a multivariate Gaussian form with larger smoothing factors.
GRNN consists of four layers: Input, pattern, summation, and output. The prediction value of feature vector $\mbf{y}$ is first computed as
\begin{equation}
    \mathrm{pred}(y)=\frac{\sum_{k=1}^{N}\mathcal{W}_k\mathrm{RBF}(y,y_k)}{\sum_{k=1}^{N}\mathrm{RBF}(y,y_k)}
\end{equation}
where $\mathcal{W}_k$ is the activation weight for the pattern layer neuron at index $k$. The radial basis function is then expressed as
\begin{equation}
    \mathrm{RBF}(y,y_k)=e^{-(y-y_k)^T(y-y_k)/2\sigma^2},
\end{equation}
where $\sigma$ is the smoothing factor.

\item{\textit{Backpropagation Neural Network: }}

Another important deep learning model affecting the robustness of the material detection and classification performance is the BPNN. BPNN is widely used to train multilayer feedforward neural networks, and it mainly consists of the input layer, one or more hidden layers, and the output layer. The BPNN principle adjusts the weight parameter controlling the degrees of connections between the neuron nodes of different layers to produce the desired output layer. Proper adjustment of weights allows minimizing the total network error. The number of input features determines the number of input neurons. Furthermore, the number of neurons in the output layer is related to the number of classes. However, the number of hidden intermediate layers between the input and output layers can be customized. 

The output of the hidden node is given by
\begin{equation}
    z_n=f\left(\sum_m w_{nm} y_m\right)= f(\mathrm{net}_n),
\end{equation}
where $w_{nm}$ is the network connection weight between the input layer and hidden layer nodes, $m$ and $n$, respectively, and where $y_m$ is the input node, and $f$ is the activation function evaluated at the sum, $\mathrm{net}_n$, produced by node $n$. The output of the output layer node is then expressed as
\begin{equation}
    o_k=f\left(\sum_n q_{kn} z_n\right)=f(\mathrm{net}_k),
\end{equation}
where $q_{kn}$ is the connection weight between the hidden and output layers.

\end{enumerate}

\section{Performance and Complexity Tradeoffs}
\label{sec:perf}

\begin{figure*}[t]
\centering
\includegraphics[width=5in]{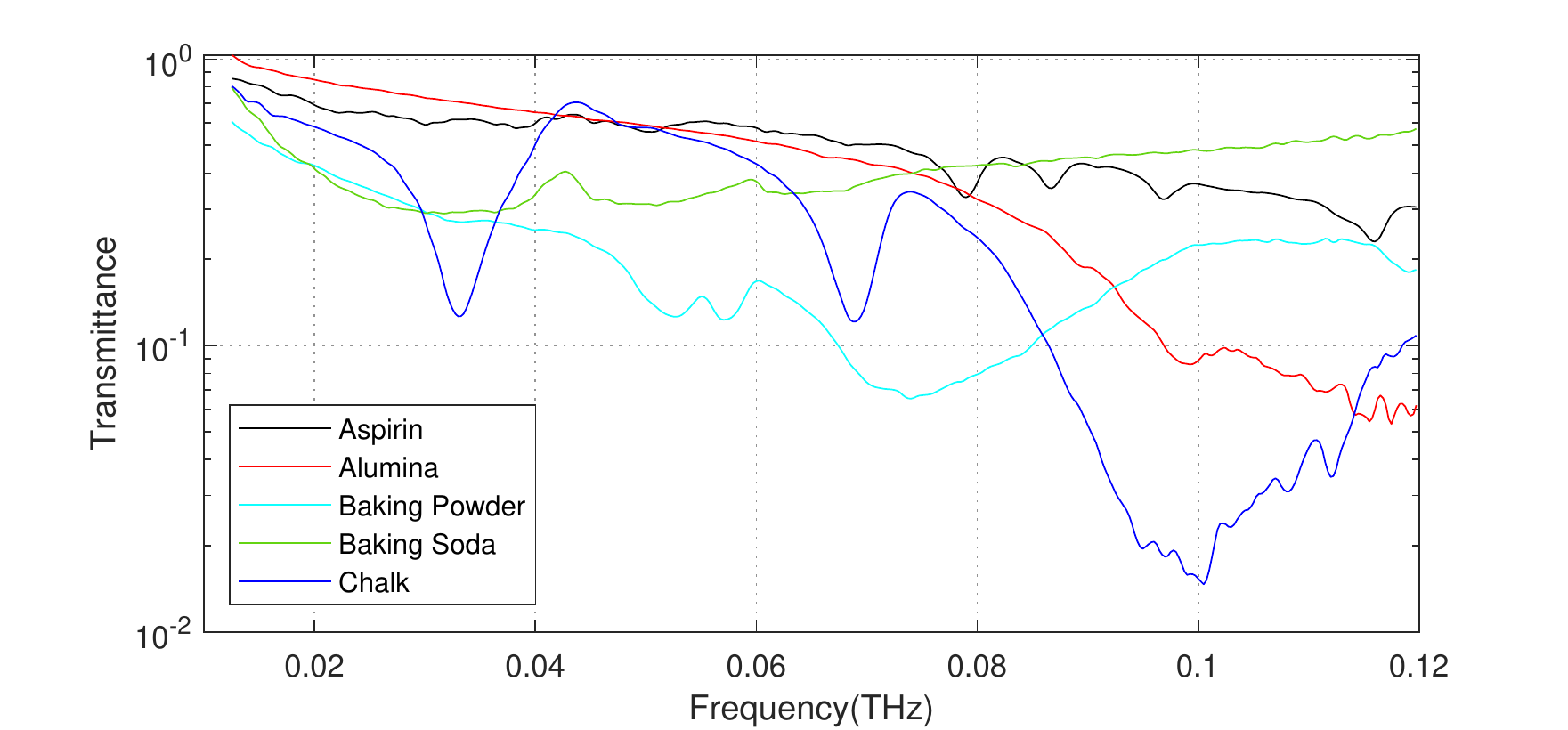}%\vspace{-0.1in}
\caption{Transmittance waveform of five materials.}
\label{matSpectra}
\end{figure*}

\begin{figure*}[t]
\centering
\includegraphics[width=5in]{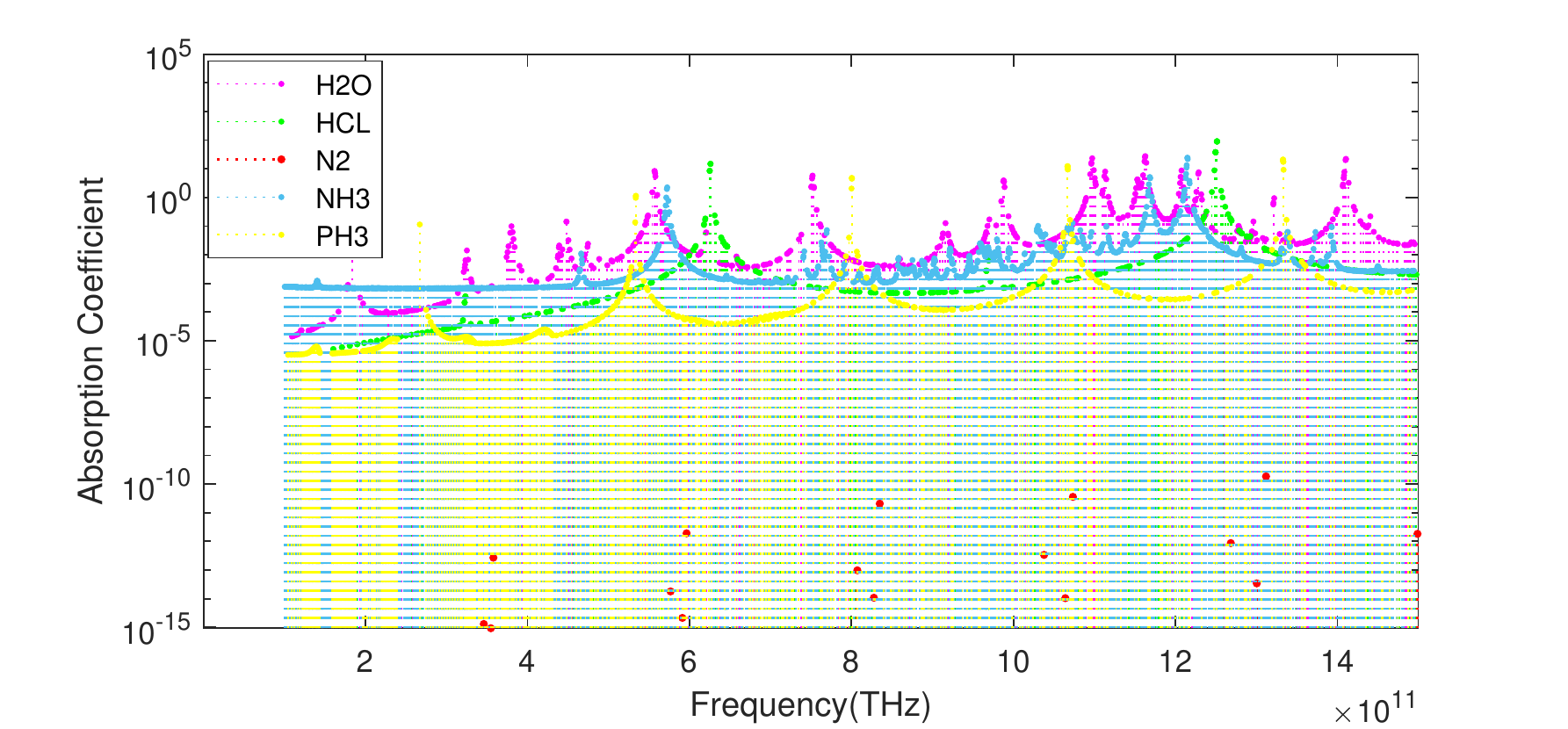}\vspace{-0.1in}
\caption{Absorption spectra of five gases.}
\label{GasSpectra}
\end{figure*}

This section analyzes the performance and complexity trade-offs of the feature extraction and classification techniques presented in the previous sections. We consider publically available THz spectral data for different materials, solids (Fig. \ref{matSpectra}), and gases (Fig. \ref{GasSpectra}), and we compare the techniques' classifications success rates.

\subsection{Terahertz spectroscopy datasets}

For material identification, we consider the sub-THz spectral data of 20 sample materials (such as alumina, aspirin, baking powder, baking soda, and chalk), as obtained from the THz database provided by the National Institute of Standards and Technology (NIST) \cite{heilweil2011thz}. The chemical materials are assumed to be grounded to a finer powder and pressed into solid pellets in polyethylene diluted pellets. By investigating the transmittance spectrum, it can be seen from Fig.\ref{matSpectra} that the THz transmittance amplitude and peak locations across the materials are different, which facilitates classification and identification. In the particular case of gas spectroscopy, the gas absorption spectra are modeled using radiative transfer theory. The molecular absorption for an individual isotope can be expressed as a function of system pressure and temperature. We extract all parameters from the HITRAN database \cite{gordon2017hitran2016}.

\subsection{Performance of feature extraction techniques }

Since smaller datasets are easier and faster to analyze and visualize than rich datasets, dimension reduction can be an essential step before implementing machine learning algorithms. For sample materials, we use 1000 observations to establish the calibration model over 430 transmission spectral coefficients (variables). We use the Savitzky-Golay function to pre-process the data for smoothing (assuming noise-corrupted data). The qualitative chemometric analysis is first performed using four different feature extraction and dimension reduction techniques: PCA, PLS regression, NMF, and t-SNE.

\begin{figure*}[t]
  \centering
  \subfloat[Relative classification performance with PCA]{\label{pca_mat} \includegraphics[width=0.5\linewidth]{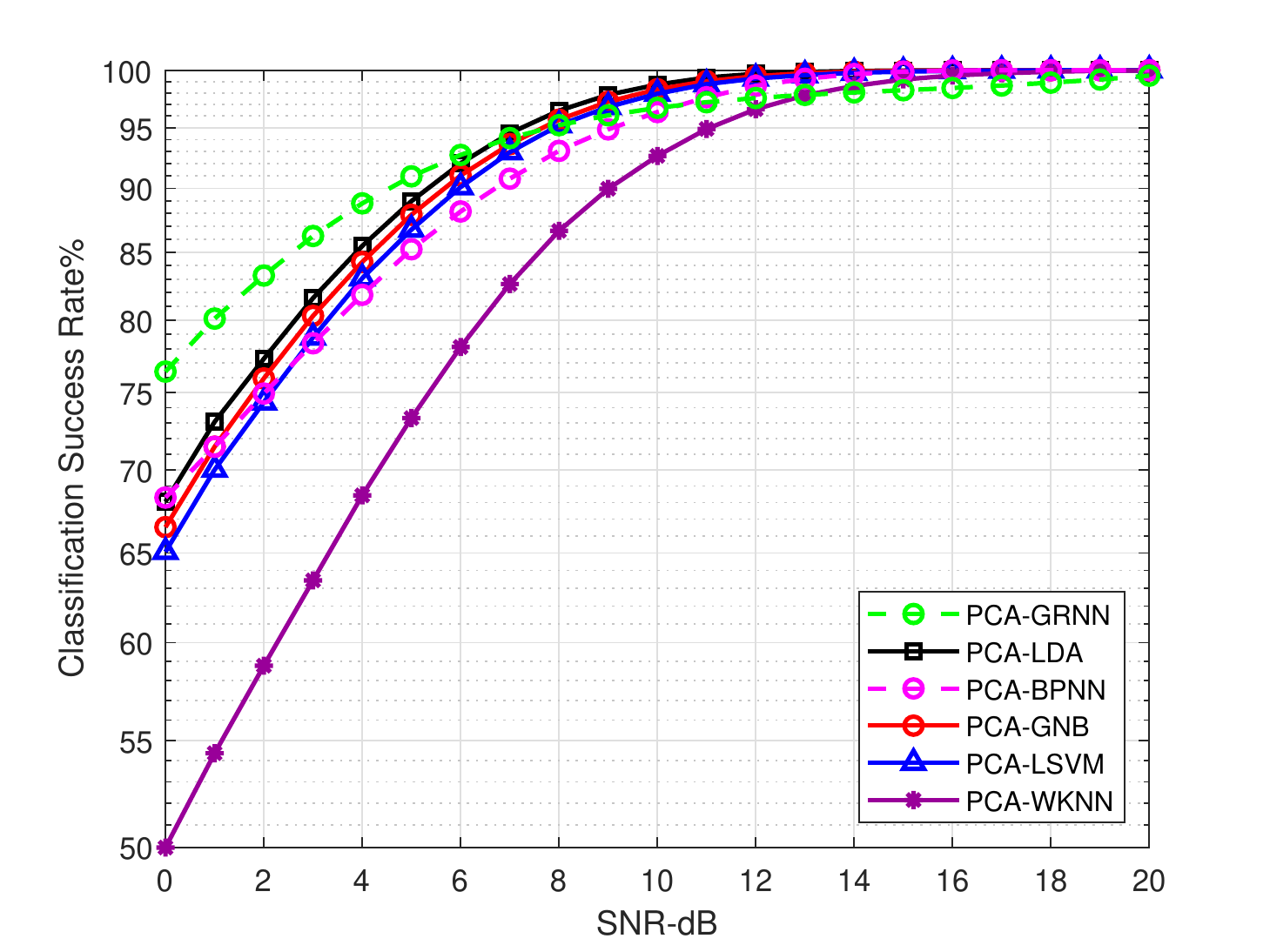}}
  %\hfillll
  \subfloat[Relative classification performance with PLS]{\label{fig} \includegraphics[width=0.5\linewidth]{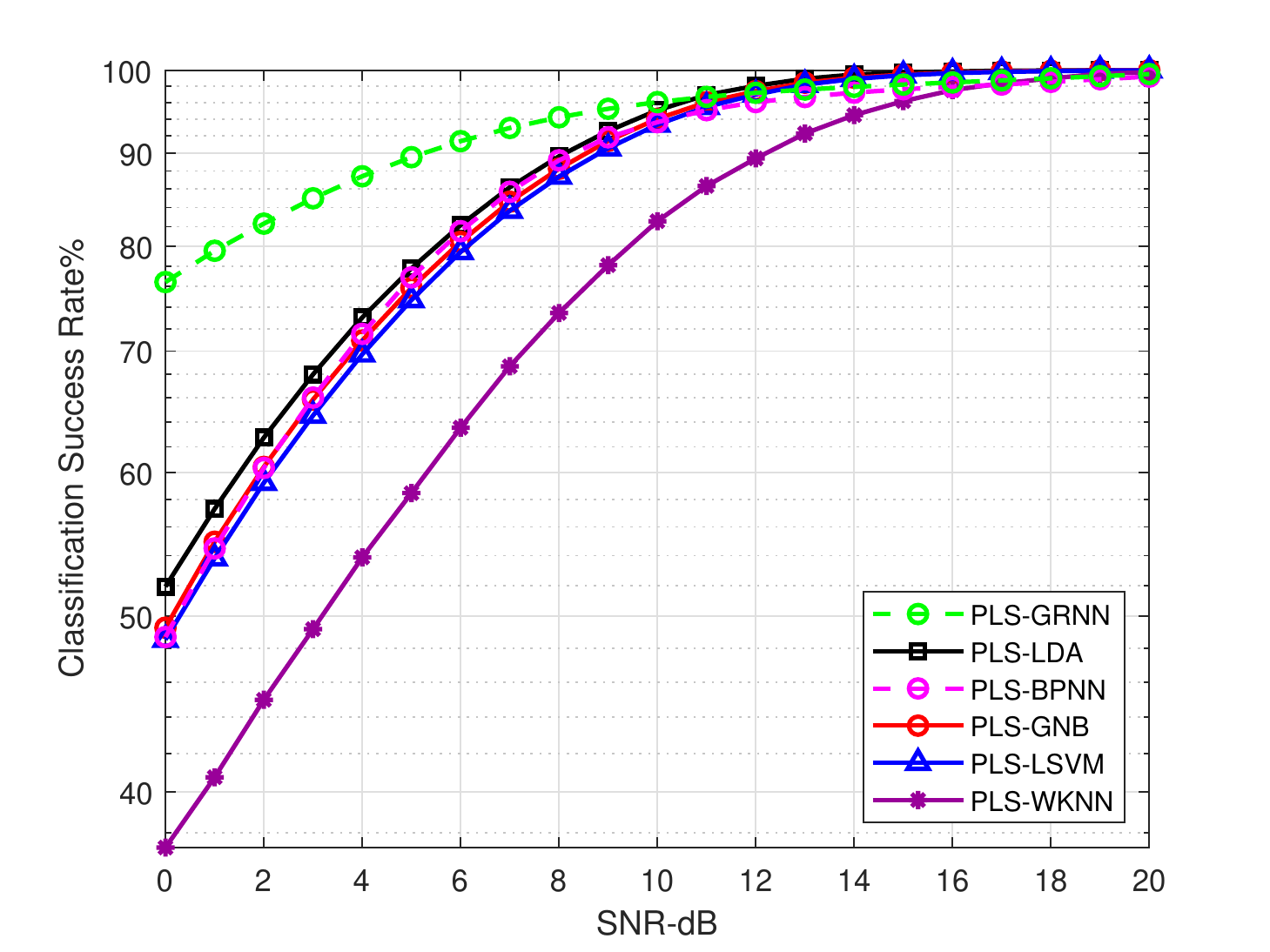}}
  \hfill

  \subfloat[Relative classification performance with t-SNE]{\label{f2} \includegraphics[width=0.5\linewidth]{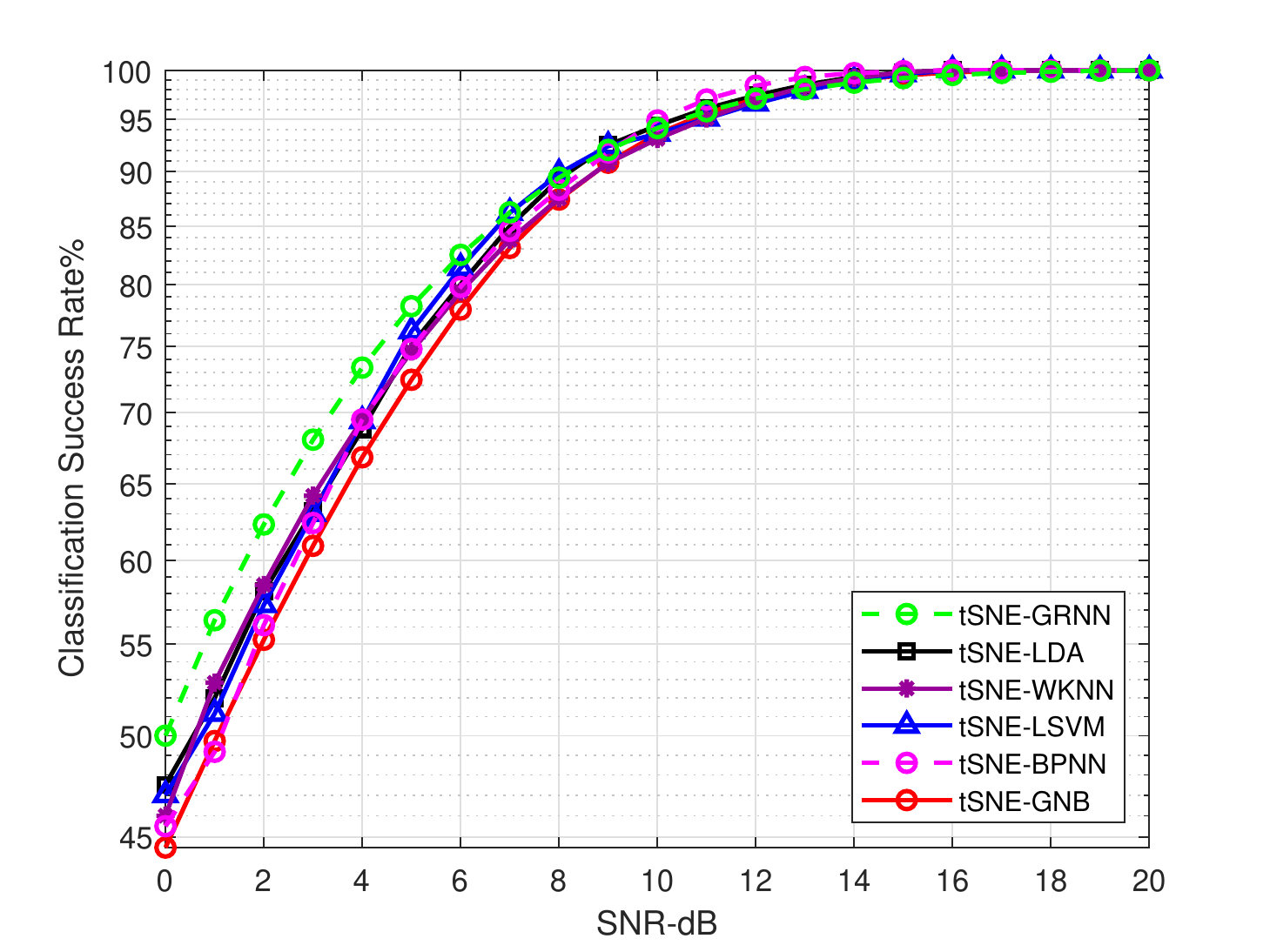}}
 % \hfill
  \subfloat[Relative classification performance with NMF]{\label{figg} \includegraphics[width=0.5\linewidth]{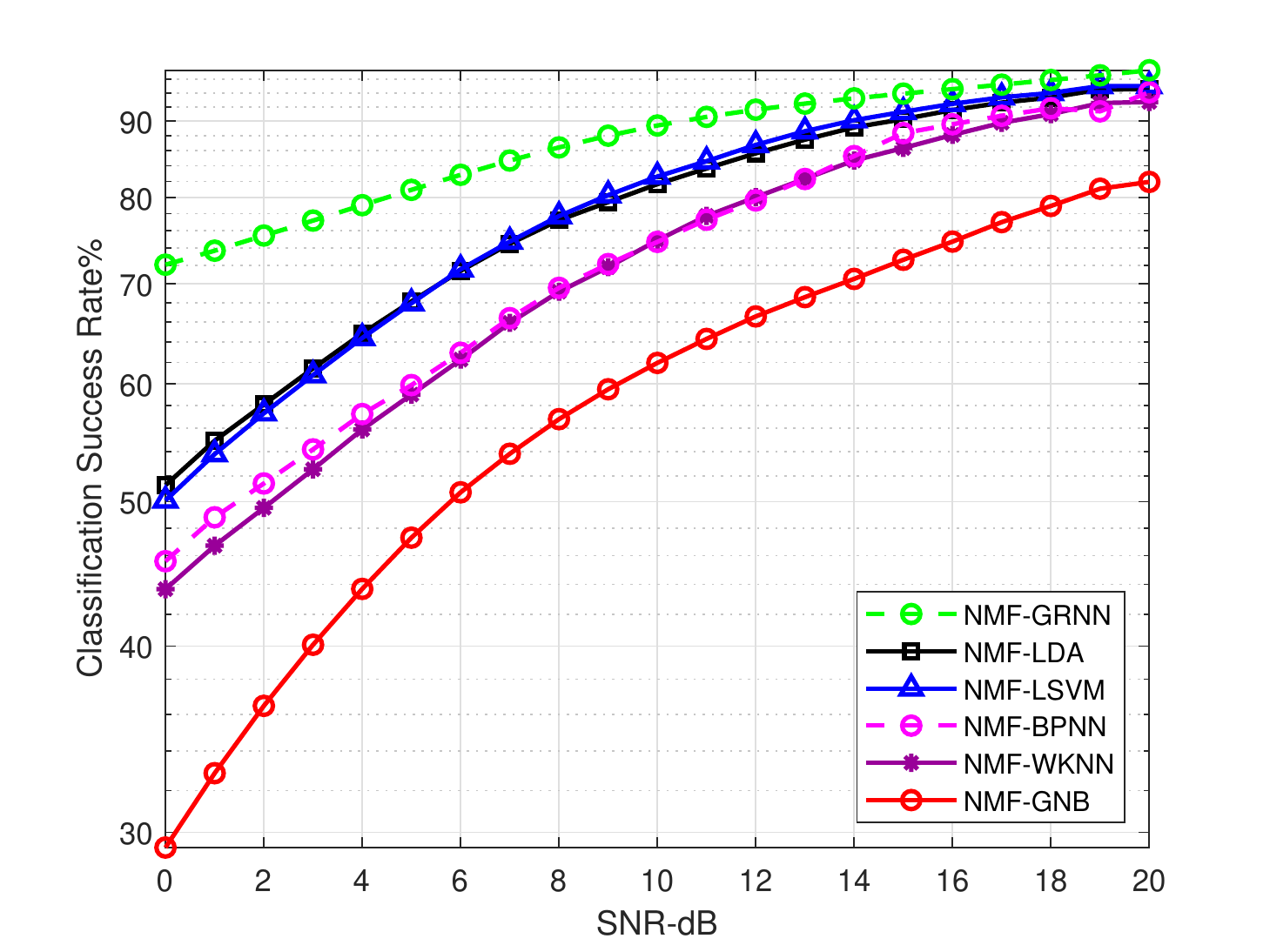}}
 % \hfill

  \caption{Performance of studied classifiers with respect to signal-to-noise ratio for solids.}
  \label{fig_solids}
\end{figure*}

\begin{figure*}[t]
  \centering
  \subfloat[Relative classification performance with PCA]{\label{pca_gas} \includegraphics[width=0.5\linewidth]{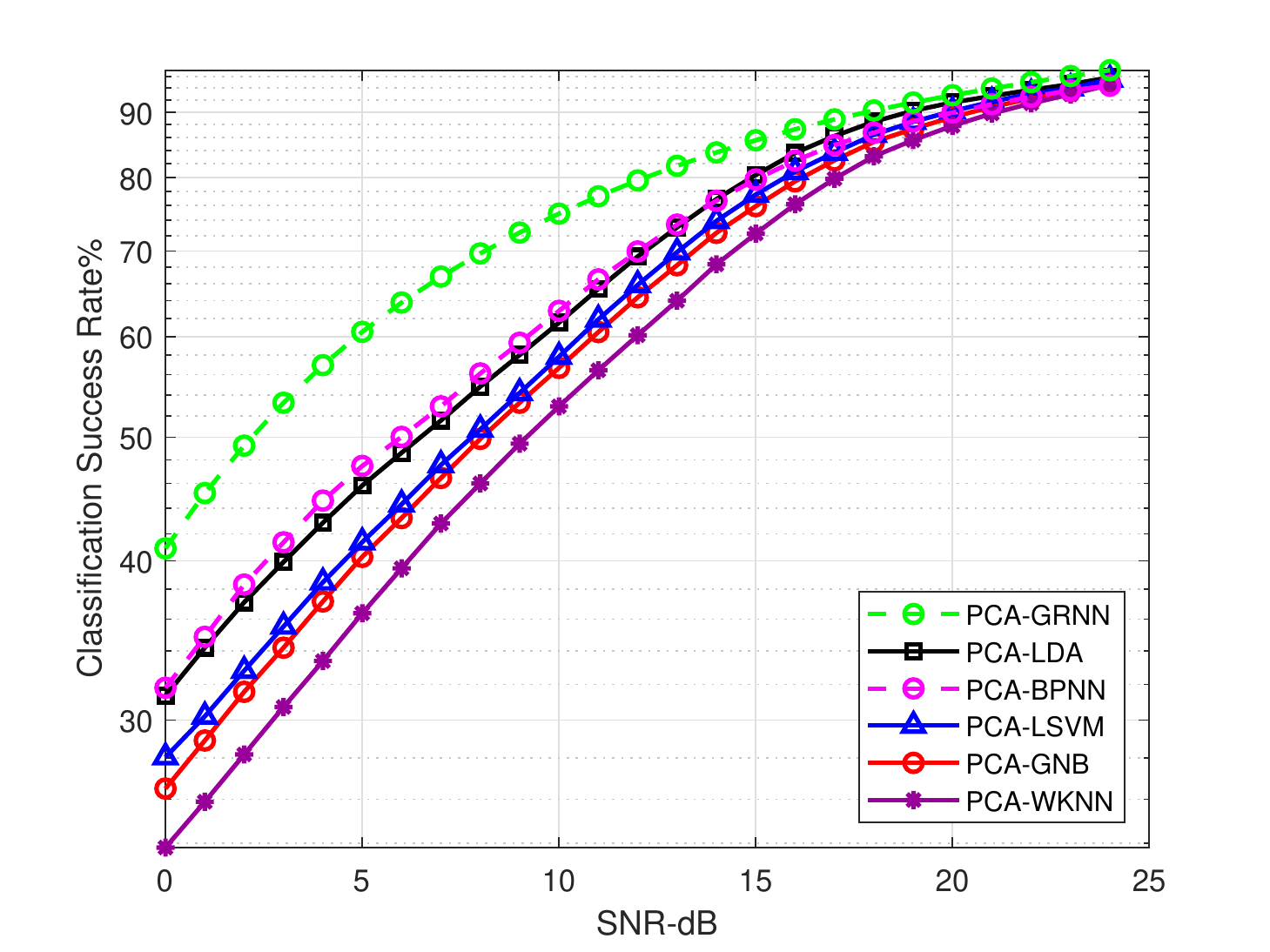}}
  \subfloat[Relative classification performance with PLS]{\label{hardfig:b} \includegraphics[width=0.5\linewidth]{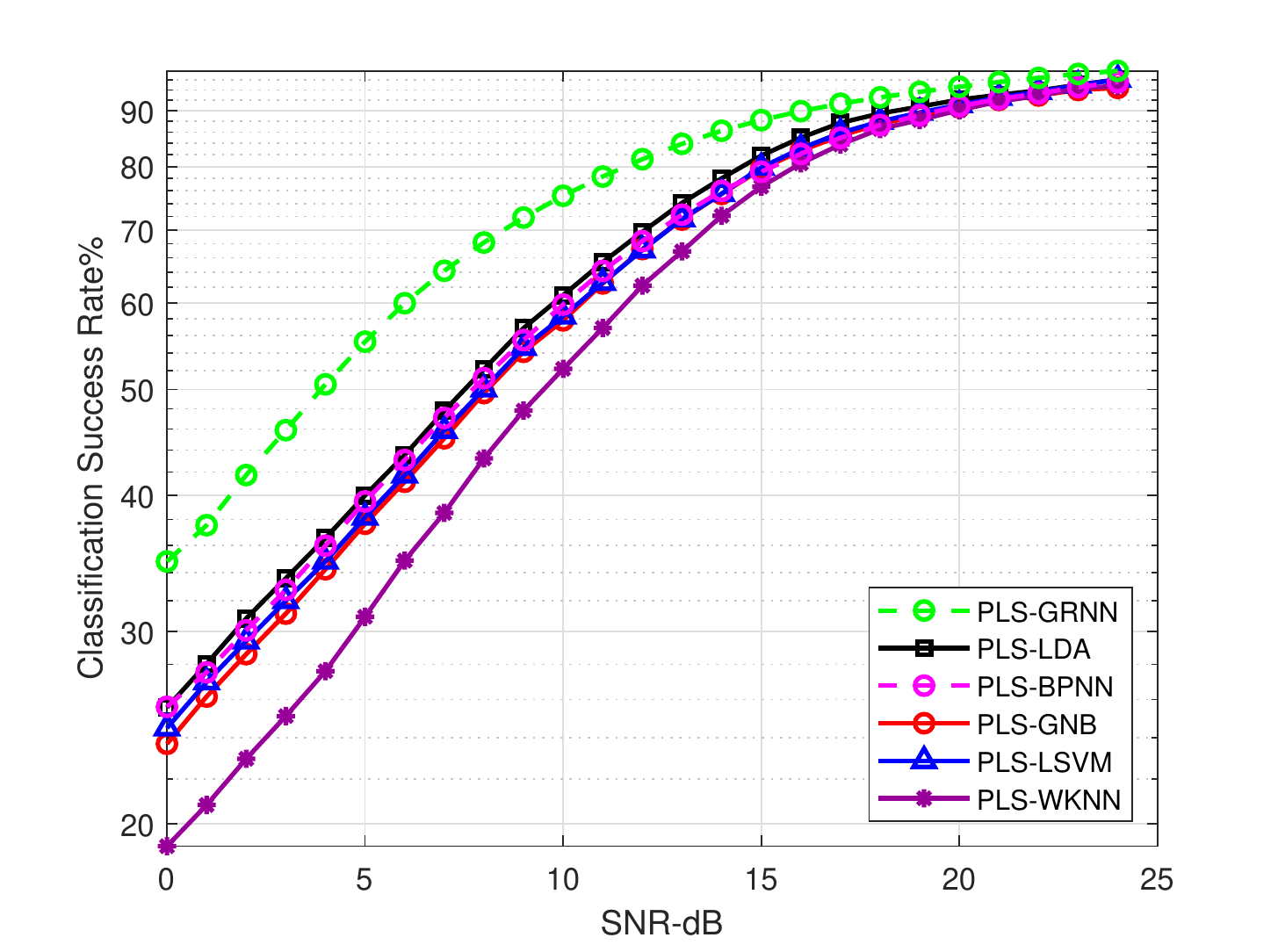}}
  \hfill

  \subfloat[Relative classification performance with t-SNE]{\label{hardfig:c} \includegraphics[width=0.5\linewidth]{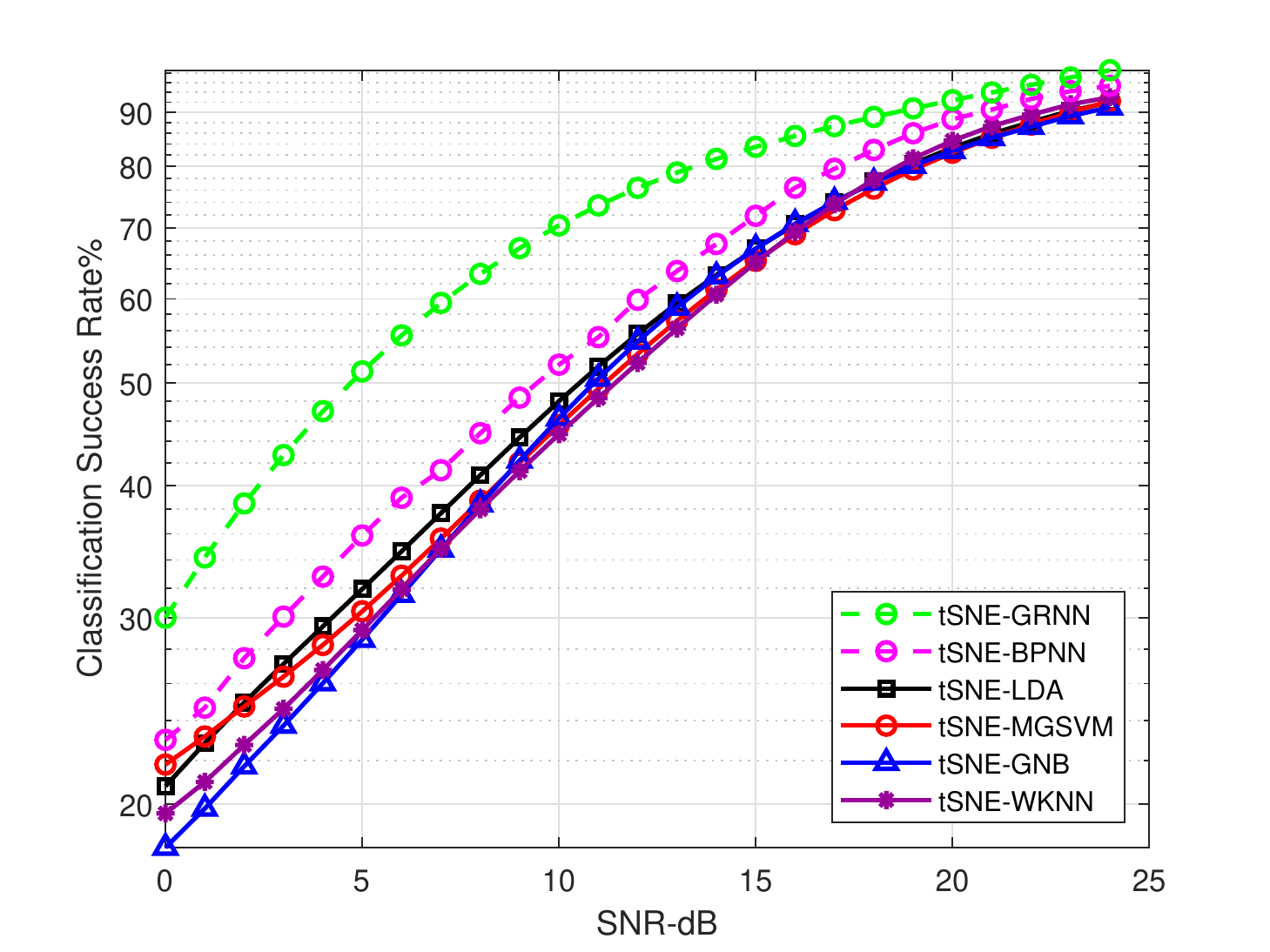}}
  \subfloat[Relative classification performance with NMF]{\label{f5} \includegraphics[width=0.5\linewidth]{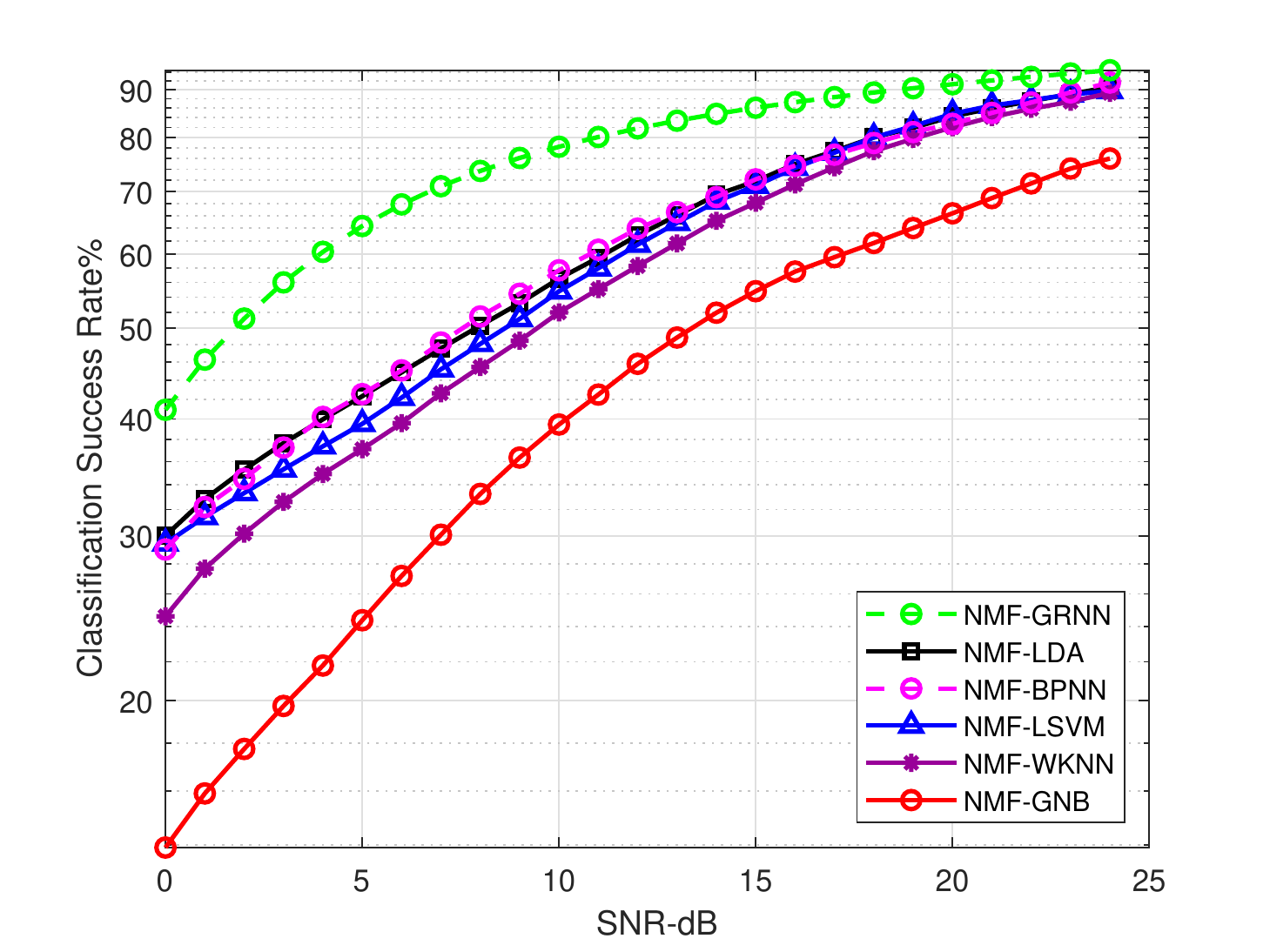}}
 % \hfill
  \caption{Performance of studied classifiers with respect to signal-to-noise ratio for gases.}
  \label{fig_gases}
\end{figure*}

For each SNR value, PCA is executed 10 times, where a small number of PCs (up to 10 PCs) is required to reach good classification performance. The THz spectral window under investigation is reduced while still retaining relevant data that captures most of the THz spectral fingerprint.
At an SNR of 20 dB, the accumulated contribution rate of the first 10 principal components of PCA reaches $97.5\%$, which results in good clustering performance. Eventually, PCA constructs PCs that convey the most variation in the available dataset. Similarly, the first 10 output components of each of the studied techniques are selected to train the classifiers. In t-SNE, we set the effective number of local neighbors of each point, known as perplexity, to 5. The best feature extraction model is the one that minimizes the computation complexity and dimensionality. We note that PCA has the best clustering efficacy (see Fig. \ref{pca_mat} and Fig. \ref{pca_gas}) for complex high-dimensional spectral datasets and that t-SNE has the highest computational and time complexities. At lower SNR, PCA outperforms NMF, PLS, and t-SNE in terms of stability, complexity, and interpretability of the spectral features. In this work, t-SNE was performed using a slightly less number of samples, as it requires considerable time to execute on the same sample size of spectral data than other feature extraction techniques.

\subsection{Performance of classification techniques }

We next illustrate the classification rate performance of several materials. We utilize linear DA (LDA), linear SVM (LSVM), weighted KNN (WKNN), and Gaussian NB (GNB) to predict the materials following each feature extraction technique. The output feature data are divided into two sets, training and testing, both of which are composed of the sample material (or class) data and a sample label.  We use a K-fold cross-validation scheme to evaluate the classification models' performance. The input data of features is partitioned into 10 equal-size sub-samples (folds), each used as a testing set and as validation data. In the first iteration, the first fold of the $K$ sub-samples is used to test the classifier, while the remaining $K\!-\!1$ folds are used to train the classifier. In the second iteration, the second fold is used as the testing set, and the rest serve as training sets. This process is repeated until each of the 10 folds is used as a testing set.

We measure the root mean square error of calibration (RMSEC) for the testing set to capture the classification model's performance. The results in Fig. \ref{fig_solids} suggest that LDA can distinguish the solid materials accurately at low SNR values with low processing time. This observation is expected because LDA is suitable for problems that deal with linearly separable THz spectral data. Gaussian NB (GNB) gave similar good classification results, which is also expected because the THz material fingerprints have continuous features and approximate Gaussian distributions. However, the GNB classifier (Fig. \ref{fig_solids}d) did not exhibit a good representation of the NMF data, mainly due to inaccuracies in its simple hypothesis function. Furthermore, the accuracy of KNN is degraded due to noisy feature data and large sample sizes, resulting in an expensive cost to calculate the distance to the nearest neighbor. The relative degradation in the accuracy of the SVM classifier, further, might be caused by overfitting. Given these erroneous classifications, the visualization of sensor data can be used to clearly show that the error in material classification is due to either of the two factors: (i) overlapping regions on the principal components space and (ii) non-correlation between data samples in training and testing datasets. The former factor is related to the chemical similarity of materials; the latter should be mainly related to the changes in the ambient conditions in the original sensor data or the drift of sensor response. In practical implementations, it can be assumed that the observed error in material classification, especially in gas identification, is mainly caused by fluctuations in the atmospheric background; some erroneous material classification can be traced as a result of sensing element degradation. To mitigate the classification errors, data augmentation, hyperparameters optimization, and more spectral data collection for the cases where errors are prominent can be some promising solutions.

For BPNN, the best performance is achieved with 10 PCs, 10 hidden nodes, and a 0.01 learning rate. However, the best GRNN performance is achieved with 5 PCs and a spread value of 10 for the radial basis function, attaining smoother function approximation. Compared to the BPNN model, the GRNN model results suggest that GRNN is more favorable in predicting the THz spectral data, mainly because GRNN has low computational complexity, intermittent flow estimation, and high computing speed. Each GRNN trains fast in one-pass learning, whereas BPNN takes much more time on average over forward and backward passes. GRNN can further converge to the THz data's underlying function with just a few training samples, unlike BPNN. Moreover, GRNN results in less classification error (better generalization ability) due to its ability to handle noise in the input data.

GRNN achieves a good balance between the high classification accuracy and speed for both solid and gaseous materials datasets (Figures \ref{fig_solids} and \ref{fig_gases}, respectively). The corresponding computational efficiency is proven in \cite{cui2020ultrafast} for all kinds of gas molecules supported by a spectral database. The results show that the multivariate discriminative model of LDA and GRNN, in association with THz spectroscopy, provides a cost-effective and low-time-consuming alternative to the commonly used models in the literature for material classification, suggesting a commercial and regulatory potential.

A closer look at the results confirms that the classification success rates largely depend on how well the theoretical basis of each classifier fits the THz spectral data. In all cases, KNN and NB classifiers have the worst performance results. Although the KNN method is conceptually appealing and computationally fast, it is not always the optimal machine learning model for remote sensing data and is sensitive to irrelevant features. As such, KNN should be merged with other techniques in this context. From Eqn. \eqref{NBeqn}, it becomes evident that the training problem in NB model is converted to a maximum likelihood one, which may lead to overfitting issues and subsequently compromises the accuracy value. However, the advantage of NB is its “naive” assumption in its algorithm that can often yield good predictive power as a result of the emphasis on evidence observed as a conditional probability of features for a particular outcome class. Moreover, one can observe a weakness in the predictive ability of SVM. This is because training SVM is computationally expensive as the number of training data increases, but at the same time, SVM has shown robustness in automatically learning data with noise and imbalanced distribution due to its superior generalization ability and the capability of dealing with linear and nonlinear problems.

In conclusion, these results confirm the advantage of incorporating the appropriate feature extraction technique based on the application served in material classification. We note the effectiveness of the generalized regression neural network approach in detecting and classifying THz sensor data compared to conventional classifiers and frequently employed backpropagation neural networks. However, it is challenging in practice to find one classifier that can classify all the spectroscopic datasets with the same accuracy.

\section{Machine-Learning-Empowered Intelligent THz Technologies}
\label{sec:MLforTHz}

Having illustrated the importance of leveraging the THz channel characteristics for material and gas sensing in the previous section, we next address the broader impact of THz technologies. In particular, we highlight several machine/deep learning applications that can empower future generations of THz communications and sensing technologies. We note that leveraging the unique THz features calls for data-driven, proactive, predictive methods capable of bypassing the highly-varying channels within the THz-band coherence times and the stringent latency requirements of 6G applications. 

\subsection{Sensing and localization} 
Most detection and recognition tasks at the THz band, such as object detection, material sensing, 3D mapping, environmental changes, and localization, measure and analyze sensing parameters from transmitted or reflected THz waves. Such measurements include the Doppler frequency, time delay, angle-of-arrival, angle-of-departure, optical constants, and other physical patterns of the detected objects. Sensing measurements can then be fed to neural networks to learn more entangled patterns and make predictive allocation of resources, in the context of localization. In conjunction with such learning mechanisms, the manifold sensing capabilities at the THz band can provide data-driven THz networks capable of achieving operational intelligence required to recognize the stochasticity of the environment in real-time. Moreover, recent advances in probabilistic machine learning and Bayesian methods, such as Gaussian processes, are bound to ultimately provide promising interpretable modeling of complex spatio-temporal and high-dimensional THz sensing problems for 6G wireless networks \cite{9782674Chen}.

\subsection{Hybridization}
The coexistence of THz frequencies with sub-6 GHz and mmWave technologies opens the door for more scalable network opportunities via heterogeneous networks that integrate a certain level of synergy between THz and lower frequency bands. When such networks start seeing the light, one major challenge is implementing novel schemes that allocate transmission interfaces between the heterogeneous frequency bands. In that sense, achieving a successful hybridization is still a rather open research direction that poses higher demand for training artificial intelligence and machine learning-based methods. Such methods should learn the most effective interface allocation while optimizing and leveraging the cooperation across the multiple frequency bands. 
 
 \subsection{Channel estimation}
 
 The acquisition of channel state information in THz systems is crucial. For example, beamforming and beam tracking need accurate channel state information to prevent misalignment issues. Although THz channels are highly dimensional and sparse in nature, THz communication systems have a limited number of radio frequency chains, making channel acquisition a challenging problem. Overcoming this challenge brings to light the role of data-driven and predictive methods capable of estimating the full THz channel state information. Different machine learning techniques have been proposed, including generative learning methods, neural networks, logistic regression, and projected gradient ascent. On the one hand, generative mechanisms are formulated, where a deep reinforcement learning framework generates synthetic environments from synthetic and existing data to train and estimate the channel under a wide range of network conditions. On the other hand, learning-based techniques are exploited to formulate a sparse recovery problem, which exploits the sparsity of the THz channel; a deep neural network estimator can acquire the sensing matrix and achieve signal recovery simultaneously. 
 
\subsection{Beam tracking} 
Utilizing THz frequencies for wireless communications in outdoor services that exhibit high mobility use cases is challenging since it requires continuous beam tracking and mobility management. However, the existing beam tracking approaches suffer from immense computational complexity incurred by the effect of wireless environment dynamics on THz transceivers. Moreover, due to the high directivity and propagation loss of THz waves, intelligent training and tracking technologies are important to obtain channel information and track mobile users. These issues call for the development of machine learning-based methods that could enhance the robust communication links performance and consequently overcome path loss and fulfill future demands. Using a sequence of previous beams and based on the beam tracking task and dataset, adaptive machine learning and regression techniques could enable the prediction of future line-of-sight (LoS) and non-LoS beams of mobile users.

\subsection{Interference mitigation}  
The uncoordinated directional transmissions by a large number of mobile users envisioned in the dense deployment of THz networks require facing challenges in inter-cell and LoS interference effects. Although massive MIMO technology has been proposed as a solution for mmWave communication networks, the applicability of such solutions is not direct for THz networks, given the poorer multi-path propagation and the lower rank channels at the THz bands. Although interference is typically received over short time intervals, the large interference power often exceeds the received power over the desired link. Machine learning and deep learning techniques can solve such challenges by efficiently estimating and mitigating intermittent interference from the directional links at a much-reduced time complexity. Therefore, exploiting intelligent methods, including reinforcement learning, is crucial to supporting massive connectivity and diverse services in THz networks.

\subsection{Open challenges}
While the aforementioned machine learning methods have been shown to achieve good performance, they also lead to multiple challenges that must be addressed. First, a large training dataset needs to be collected for efficient database creation, which is not affordable and calls for more research on data augmentation suitable for the THz band. Moreover, the scalability challenge of the learning methods remains partially tackled. The latter is true because the collected data for model training are scenario-dependent, leading to a continuous update in the THz data in reference to the environment and system in use \cite{9782674Chen}. To overcome this issue, intermediate sensing parameters such as the angle-of-arrival can first be acquired utilizing learning-enabled methods instead of directly acquiring an end-to-end position estimate. Such solutions enable intelligent systems suitable for THz sensing and localization to adapt to various scenarios and environmental variations, reducing computational complexity considerably. Another challenge worth mentioning is the long training periods that obstruct current learning systems from operating effectively in real-time. Offline design and training, where no prior knowledge of the sensing/localization data is available, might not be an appropriate solution, given THz signals' non-stationarity and peculiar specialties. A promising alternative is, therefore, is to develop real-time machine learning techniques that incur shorter training periods and improve the training process.

\section{Joint THz Communications and Sensing}
\label{sec:JCAS}

The progress made on THz system design triggers a plethora of promising research directions, especially on the convergence of communications and sensing. To this end, we next present some specific, timely applications of joint communications and sensing at the THz band, highlighting the underlying system design considerations.

\subsection{Joint THz Communications and Sensing Applications}

The prospective use cases of THz sensing are mainly in the context of joint communications and sensing \cite{sarieddeen2019next,articleChaccour}. A unified THz system for communication and sensing can support various applications, from in-home digital health to building analytics (e.g., residential security). For instance, the THz band can establish reliable communication links for unmanned automotive vehicles (UAVs) that build on accurate localization and sensing capabilities. The THz band can also provide high-rate virtual reality services, enabling good visual perception. In particular, THz signals can support extended reality (XR) interfaces capable of interacting with sensory information in indoor environments without intervention \cite{articleChaccour, inproceedingsChaccour}. Sub-THz vehicular communications and sensing can further enable data exchange in vehicular-to-everything (V2X) communication systems. The deployment of such unified systems requires exceptionally high data rates, low latencies, and high reliability. Several contemporary approaches to joint vehicular communication and sensing are explored, such as time-domain duplex (TDD), telecom messages over radar transmissions (ToR), and radar sensing over telecom transmissions (RoT)\cite{8722599Petrov}. Moreover, joint THz communications and sensing applications are particularly useful in body-centric systems where THz-TDS can be used to sense infections and mitigate virus outbreaks\cite{9650789Saeed}.

\subsection{Machine Learning for Joint THz Communications and Sensing}
Intelligent THz systems should achieve high-rate communications and robust high-resolution sensing at the same time. Towards this end, efficient signal processing techniques are critical. Recent advances in machine learning and compressive sensing can significantly accelerate THz network applications by enhancing situational awareness and facilitating fast and low latency configurations. In this regard, network intelligentization is a new trend that aims at leveraging machine learning techniques to empower 6G communication systems with artificial intelligence (AI) algorithms. However, unleashing machine learning's full potential requires addressing significant challenges, especially in waveform design. The differences in performance metrics between communication and sensing functionalities (achievable data rates, level of interference, sensing accuracy, and reliability) should also be taken into consideration. For instance, deep learning and probabilistic methods can be applied for target classification, waveform recognition, material sensing, and optimal selection of antennas and radio frequency chains in a JCAS setup. In data-demanding and complex JCAS applications such as global navigation satellite system (GNSS)-based applications, where poor channel conditions exist in both indoor and outdoor urban scenarios, machine learning techniques are essential to realizing high-dimensional, multi-modal, indirect, and noisy spectral fingerprints and modeling the physical characteristics of nonlinear THz signals. Furthermore, it is essential to characterize how such joint and mapping systems behave in several cases, e.g., the sensor noise and uncertainties of THz systems. Such a process would likely be achieved by hybrid AI methods which are able to learn from the combination of conventional physics-based models of signal propagation and machine learning with sequential Bayesian learning in state-space models \cite{Lima9330512}.

\subsection{Network Architectures}
\begin{itemize}
\item \textit{Heterogeneous networks:} Incorporating heterogeneous networks comprising THz, lower frequency bands, and optical wireless communications allows next-generation wireless systems to achieve energy-efficient universal coverage, improve link reliability, and provide more versatile functionalities. For example, from a JCAS standpoint, the synergy of THz with mmWave frequencies can exploit sensing services for highly mobile user equipment and outdoor applications such as radar and environment sensing. This dual-sensing can further improve situational awareness and blockage mitigation \cite{articleChaccour} which, however, gives rise to complexity in deployment, and a tradeoff between the spatial resolution attained and the communication distance. For instance, THz sensing achieves high precision in detecting objects within short distances, while mmWave sensing can better detect distant objects but with lower precision. Therefore, it is necessary to have continuous feedback between the sensing and communication signals, diminishing the uncertainties surrounding the high-frequency bands \cite{articleChaccour}. 
\item 	\textit{RIS-assisted networks:} For most 6G applications, THz-band signal propagation distance and coverage range are limited. However, the rise of reconfigurable intelligent surfaces (RIS) provides a potential solution to compensate for propagation losses and creates value propositions for the 6G technology revolution. For example, the envisioned features of RIS deployment promise holographic communications (due to short THz wavelengths), near-field communication capabilities, high-flexibility, low-complexity deployments, and reshaped wireless transmission environments. RISs are equipped with massive sensing elements that constitute an intelligent sensing platform for environment sensing, object detection, and health monitoring purposes, leading to wireless intelligence. Connected with narrowband THz radiation, multiple RISs can be exploited for sensing \cite{Lima9330512}.

Moreover, using RIS-assisted architectures can improve the joint communication, sensing, and localization performance by enabling a massive number of independent propagation paths to enhance THz-based detection and localization of people and objects within a network. This is possible by optimizing the programmability and reconfigurability of  RIS meta-surfaces. Depending on the operating modes of RISs, for example, intelligent reflectors or transceivers, the performance can be limited by the tradeoff between energy efficiency and signal processing capability. Here, the RIS is an intelligent reflector that can achieve high energy efficiency for the network operator's sensing and communication control. Note that RISs exhibit low processing and transmission capability. Hence, whenever deployed as transceivers, RISs can improve performance in terms of transmitting and processing sensing and communication electromagnetic radiations, but at reduced energy efficiency and increased cost and infrastructure complexity \cite{articleChaccour}. Therefore, RISs can extend the gas and material sensing range (seeing behind corners, for example), but the effect of the intelligent material itself on the reflection coefficient should be accounted for.

\item \textit{Cell-free massive MIMO network:} UM-MIMO technology can potentially provide 5G and beyond THz systems with leveraged pencil beamforming gains, spectral efficiency, multiplexing benefits, and enhanced achievable rates to combat the propagation distance limitation in THz systems. Moreover, adopting an array-of-subarrays (AoSA) configuration allows further improvement in such gains and reduces hardware costs and power limitations by enabling hybrid beamforming in both digital and analog domains for THz communications. Given the challenges of deploying dense UM-MIMO systems, massive cell-free MIMO is envisioned as a key enabler network technology that is particularly beneficial for THz sensing and communication systems. This network architecture reduces handovers and the likelihood of handover failures resulting from the dense deployment of THz networks and diminishes LoS interference. Furthermore, by adopting cell-free networks, each user equipment could have high coverage probability, and clustering diversity of multiple base stations to a given user can also enhance the sensing, communication, and localization performance. 
\end{itemize}

\subsection{Hardware and beamforming design}
Given that the hardware is jointly shared between sensing and communication, the design of such systems requires low-cost, compact-size transmitters and wireless sensor networks from the 6G outset. This necessitates the joint hardware platform to ensure less power consumption, spectrum sharing, and enhancement in performance and safety due to sensitive information sharing. Moreover, directionality selection of antennas and quantization effects of phase shifters on antenna arrays are critical design considerations for THz system deployment. The use of directional antennas is generally required, and, in turn, this enables very directional sensing, localization, and imaging at much finer spatial resolution. In the context of JCAS, real-time network control needs to be taken into account to meet the conflictive requirements for the sharply pointing and highly directional beams for communication and the varying directional scanning beams for wide-range sensing.

\subsection{Opportunities and Challenges }

Compared to disjoint systems, future joint sensing and communication systems present many promising opportunities and advantages. The key advantage of such versatile THz systems is reduced cost and size. Without requiring major changes to the communication functionalities, sensing processing techniques can be conducted over the received signals. Moreover, JCAS systems can pave the way for enhanced beamforming and beam tracking and improved user association, enabling communications served by the sensing feedback links. Deploying joint systems allows sharing the allocation of spatial resources such as the number of antennas and metasurfaces and increases spectrum efficiency and sharing under broad THz bandwidths. In addition, THz joint systems can serve applications in indoor areas, where instantaneous localization and high data rates are necessary. 

One can benefit from JCAS systems for many new use cases, such as extended realty services, Internet of Everything, and wireless 6G systems, that require mutual sensing and communication feedback. Target applications of machine-learning-aided sensing and JCAS/localization at the THz band can hence range from low-level feature extraction and pattern recognition to object detection, position tracking and prediction, wireless channel charting, and automatic environmental mapping, to name a few. However, the features of such technologies in future THz systems will be more autonomous, time-evolving, and non-stationary, motivating the use of online adaptive machine learning methods.

The real-world deployment of THz systems that provide joint communications and sensing capabilities encounters many challenges in network design, modeling, and analysis. First, the short communication range and intermittent nature of THz links limit the reliability and coverage of communications, the sensing scope, and the situational awareness coverage. Second, although JCAS systems allow both static and dynamic resource sharing and allocation, they give rise to a tradeoff between the sensing resolution and the high data rate in dynamic resource allocation. Third, a higher transceiver complexity of the joint systems is caused by differences in waveforms designated for radar sensing and communication signals \cite{articleChaccour}. Also, deploying dual-frequency band systems (for example, symbiotic integration of THz with mmWave) to long-rang applications leads to problems in the feedback mechanisms between sensing and communications. Therefore, it is necessary to take into account these issues to achieve real-time network optimization and design novel features for the THz joint systems.

\subsection{The Role of THz Frequency-Domain Spectroscopy}

The THz band provides the possibility of designing both carrier-based and pulse-based setups that offer robust wireless sensing functionalities in communication system frameworks. However, transmitting short-time THz pulses in THz-TDS that cover large THz frequency bands is inefficient for communications. A carrier-based THz sensing setup can offer better flexibility in jointly meeting the continuously increasing bandwidth demands and sensing capabilities. We denote by the latter THz Frequency-Domain Spectroscopy (THz-FDS). THz-FDS can be conducted to recover the channel's molecular, spatial, and temporal characteristics when probing the environment with a discrete set of carrier frequencies. THz-FDS wireless sensing comprises three steps: (1) Selectively probing few THz waves into a medium, (2) estimating the channel response to identify THz fingerprints, and (3) correlating the estimated response with a reference spectral database of the constituents of the medium under investigation. 

However, the spectral lines that form fingerprints for testing occur at specific resonance frequencies that are typically avoided when allocating carriers for communication purposes. On the one hand, this observation complicates the joint communications and sensing problem because near-absorption-free communication spectra hold little sensing information. On the other hand, if sensing is not fully piggybacked over communication resources and time-sharing is enabled, the carriers can be directly tuned to the target resonant frequencies of specific gases/materials to achieve better accuracy.

\begin{figure}[t]
\centering
\includegraphics[width=3.5in]{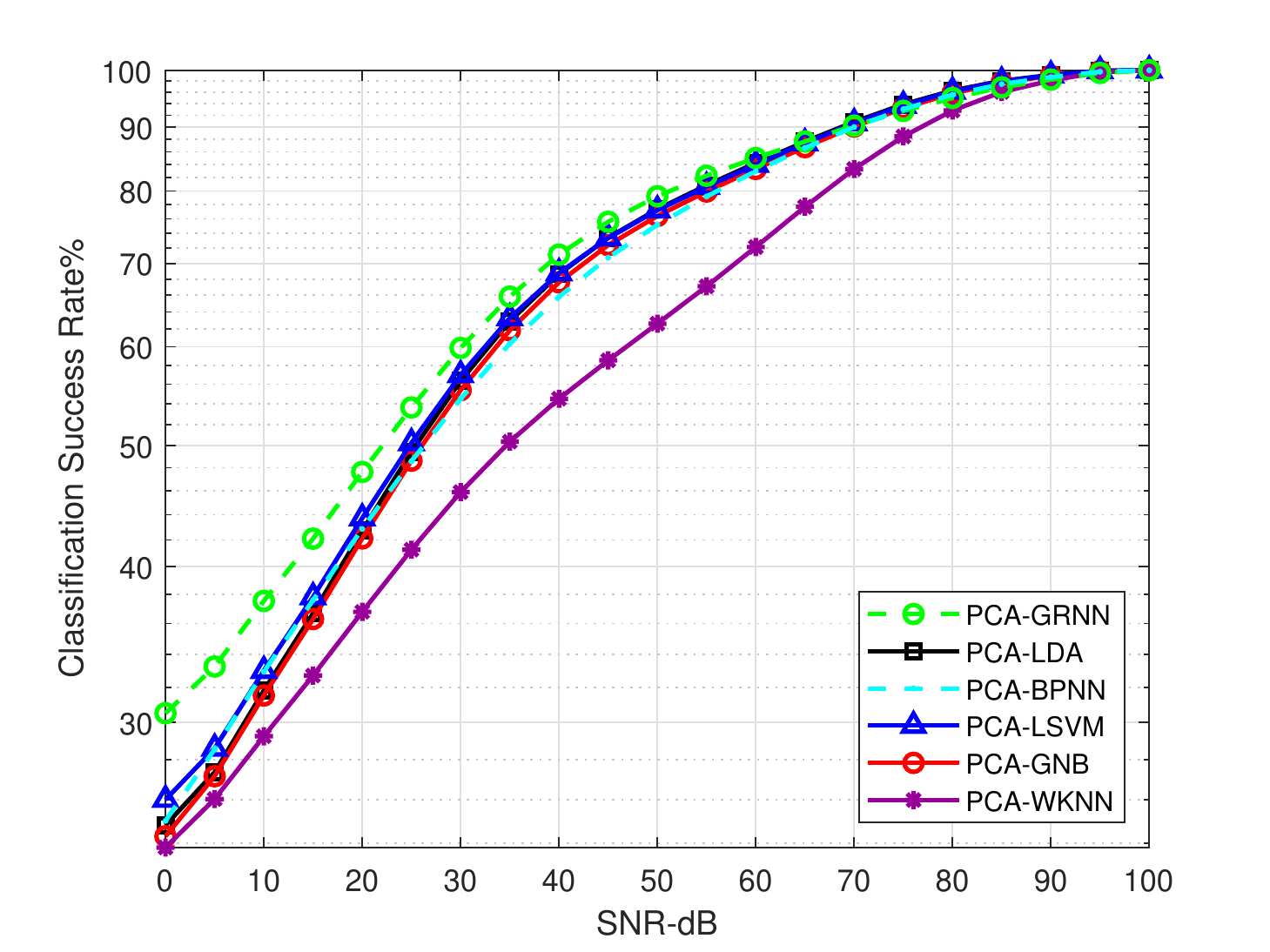}\vspace{-0.1in}
\caption{Performance of studied classifiers with THz-FDS sensing of a mixture of gases.}
\label{f:joint_gas}
\end{figure}

\begin{figure}[t]
\centering
\includegraphics[width=3.5in]{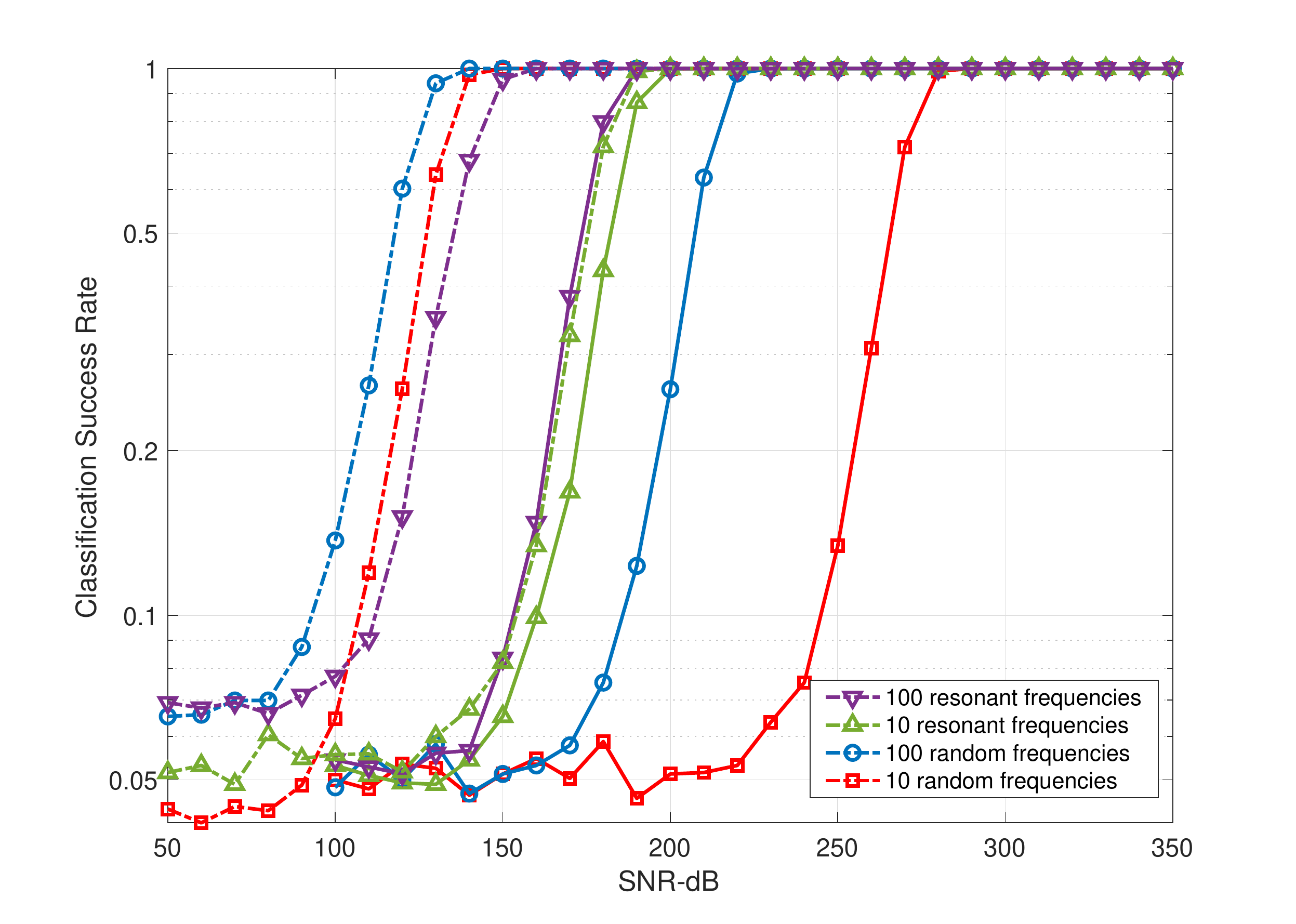}\vspace{-0.1in}
\caption{Performance of carrier-based joint communications and sensing of O2 (solid) and H$_2$O (dotted).}
\label{HeuApp}
\end{figure}

We test THz-FDS in the context of environment electronic smelling (gas sensing). We assume several realistic mixtures of gases (N2, O2, H2O, CO2, and CH4), demonstrating different possible mediums with molecular concentrations representing dry, humid, and polluted air profiles. The corresponding results in Fig. \ref{f:joint_gas} demonstrate that carrier-based sensing of gas mixtures is feasible. However, higher SNR values are required for convergence to a 100\% classification success rate because estimating a mixture of gases is complicated. Not that we used 100 carriers, randomly distributed over a specific frequency range, which can be allocated in a single channel use with orthogonal frequency-division multiplexing (OFDM), for example, given the large THz bandwidths. We also test simple heuristic search algorithms to illustrate the importance of tuning carriers to resonant frequencies in joint communication and sensing setups, as illustrated in Fig.~\ref{HeuApp}. Tuning 10 or 100 carriers to resonant frequencies of water vapor or oxygen introduces significant gains compared to uniformly distributing these carriers between $\unit[1]{THz}$ and $\unit[1]{THz}$. Note that the high SNR values are caused by the high propagation losses of THz channels (over a distance of $\unit[5]{m}$) that are accounted for in this simulation. Lower SNR ranges can be achieved by adding substantial antenna and beamforming gains.

Towards introducing beamforming gains, UM-MIMO systems can be deployed. High beamforming gains increase the received signal power and provide the required high-resolution spatial focusing at a specific distance (molecular absorption is also distance-dependent). Furthermore, UM-MIMO systems can realize multiple measurements in a single channel use. However, the high correlation between absorption spectra and the inherent high correlation of UM-MIMO channels results in low-rank measurement matrices. Spatial tuning of antenna separations can guarantee orthogonality of THz channels to achieve high multiplexing gains \cite{9634116Hadi}. The level of accuracy also depends on the application requirements and the assumptions on how many gases and isotopes of gases can exist in a medium. Therefore, the problems under study can easily get prohibitively complex for simple signal processing techniques, hence the motivation for machine learning. 

Instead of comparing the exact values of channel measurements, we can set thresholds to check the presence or absence of specific spikes and build decision trees for classification \cite{ryniec2012terahertz}. Furthermore, when a small number of gases/materials is being tested, the corresponding sparsity can be leveraged in compressive sensing techniques. Note that the transmitted information-bearing symbols over the channel can be assumed to be random for sensing purposes. However, in cooperative sensing and communications setups, such symbols would belong to a specific constellation with a specific structure, a quadrature amplitude modulation, for example. The knowledge on modulation format can further be exploited to enhance the sensing performance. Note that in adaptive THz UM-MIMO systems, the set of active antennas, the carrier frequencies, and modulation modes can be tuned in real-time while being efficiently blindly estimated at the receiver side \cite{loukil2019terahertz}.

\section{Future Research Directions}\label{future:res}

Shifting to THz technology indeed calls for better sensing and localization performance. However, new challenges need to be considered to successfully deploy and operate the next-generation THz systems. This section suggests some future research directions which are expected to be at the forefront of THz sensing and communications studies:
\begin{enumerate}
\item{\textit{Towards advanced machine learning: }}
We envision that high user quality of experience (QoE) can be achieved by migrating towards cross-layer intelligent THz systems, capable of addressing THz channels' peculiar properties and uncertainties, in which sensing data and network communication data are fed to real-time and out of the box machine learning techniques.
\begin{itemize}
\item{\textit{Multi-task learning: }}
Given the susceptibility of THz signals to extreme events of blockage and deep fades, standard machine learning approaches like reinforcement learning might not be capable of finding reasonable and accurate solutions when encountered with out-of-distribution data that are far away from the training distribution learned. In that sense, one could implement multi-task learning techniques, which have shown promising results, by tackling multiple learning tasks simultaneously instead of learning single tasks. As such, machine learning-enabled THz systems can capture an umbrella distribution that entails all the possible outcomes by jointly learning multiple tasks, enhancing learning agents' scalability.

\item{\textit{Meta-learning:  }}
Given the low-latency requirement of 6G services, dealing with reinforcement learning algorithms is troublesome due to their long training periods that could subsequently disrupt the real-time operation of such services. To overcome this challenge, meta-learning solutions help mitigate the training and exploration periods of reinforcement learning schemes \cite{articleChaccour}. Thus, meta-learning is indeed an effective approach that acquires fast adaptation to dynamic and non-stationary environments, leveraging the machine learning methods for both wireless and non-wireless settings. However, meta-learning is still in its early stage and a rather open research direction to be investigated in the realm of 6G technologies.

\item{\textit{Deep learning: }}
The general need for automating analytical modeling, empowered by the developments in machine learning techniques, is trending nowadays, both from numerical and theoretical perspectives. Since at the core of the THz sensing problem lies a pattern recognition procedure and, by transitivity, a designated function approximator (as entailed earlier in this paper), we expect the dynamic machine learning developments to play a major role in future THz sensing research directions. For instance, in sensing applications characterized with long-term variable dependencies, one can investigate the merits of using convolutional neural networks, recurrent neural networks, or echo-state networks \cite{ESNcoh2} as potential sensing solutions. In sensing problems where distributed, yet privacy-preserving, implementation is of practical importance, we can adopt federated learning-based sensing techniques. Lastly, in problems involving interactive learning frameworks, e.g., in sensing situations where one agent would be able to sense one state of the environment and consequently executes an action that impacts the next state for maximizing a certain score (reward), one can use reinforcement learning-based techniques. Such a reinforcement learning paradigm would indeed be quite fitting for THz multi-purpose platforms, highlighting the THz band capabilities as a powerful enabler of joint communications, sensing, and localization, which promises to be an exciting area of future research in the field. 
\item{\textit{Active learning:  }}
In parallel with advances in machine learning techniques for THz applications, active learning becomes a natural choice for a multitude of remote sensing applications, including the detection of defects in real industries. Due to the expensive labeling efforts, the deployment of object detection and localization systems into more complex and challenging real-world environments is difficult. Given the stochastic problems of wireless networking and lack of labels, a deep reinforcement learning framework needs to be re-architected. Active learning could be used to reduce labeling efforts and reach optimal performance.

\item{\textit{Specialized learning:  }}
Future research on machine learning-based THz applications will focus on empowering THz systems with specialized machine learning models that are best suited and designed for a particular task \cite{Parks21041186}. For instance, if the target sensing data is disrupted by attenuation, noise, or other interfering factors, which is a highly relevant case for THz spectral data obtained from liquids and biomolecules, applying simple machine learning techniques cannot guarantee a good performance. The first step towards designing a specialized model for a given task in the context of sensing is to analyze the characteristics and constraints of the signals pertaining to the THz system in use and the unique properties of test samples. The final model can then be designed and optimized for each task by investigating how they influence learning behavior in different machine learning models. 
\end{itemize}

\item{\textit{Towards ubiquitous coverage:  }}
Transitioning towards real-world deployment of JCAS systems at THz requires an evolution in the wireless systems for well-designed THz architectures to account for the short communication range and highly varying and uncertain channels of THz links. Subsequently, this necessitates new methods to maximize and achieve maximum coverage of THz networks. In the context of smart radio propagation environments, one potential solution to the coverage challenge is to exploit deep reinforcement learning (DRL) \cite{Huang9367503}, which leverages the advantages of deep learning and iterative updating through interaction with the environment. Several recent research interests and advances uncover DRL-enabled technology to be used in the future 6G wireless communications, combatting the severe propagation losses and optimizing the coverage range at THz-band frequencies. However, this scheme has latency-sensitive deep neural networks, which calls for training latency-reducing methods.

\item{\textit{Towards holographic design:  }}
By investing in intelligent surfaces and holographic metasurfaces, THz systems can enable better sensing and more reliable joint sensing and communication functionalities, which are essential processes for improving wireless localization, radar, and situational awareness simultaneously \cite{articleChaccour}. In light of this, the possibility of identifying and tracking the objects in an autonomous cyber physical system is expected to go hand-in-hand with the deployment of THz sensing system, enabling diverse highly challenging use cases, e.g. advanced health monitoring and control of industrial processes that need atomic and molecular precision. Not surprisingly, the safety-critical nature of such systems rely heavily on machine learning image classifiers to provide adaptive sensing and imaging interfaces. Moreover, such design holds an effective solution for high data transfer rate and connection reliability for THz wireless local access application scenarios, such as high-definition holographic video conferencing. Specifically, nonparametric Bayesian methods and Q-learning-based approaches can be implemented in proximity-based authentication and access control to detect the devices present in the proximity while preventing the leakage of localization and other sensitive data of the devices \cite{Boulogeorgos}.

\item{\textit{Addressing health and privacy concerns:  }}

According to the International Commission on Non-Ionizing Radiation Protection (ICNIRP), the predominating thermal effect of THz radiation is considered a health risk factor in wireless systems. Unlike other radiations such as X-ray, THz is non-ionizing. Furthermore, THz signals cannot penetrate the body beyond the skin. As such, the main health risk is confined to heating of the eyes and skin caused by the absorption of THz energy in body tissues. Nonetheless, whether THz radiation demonstrates adverse health effects, e.g., skin cancer, is still not clear. Given the enormous network densification with massive MIMO and high antenna gains at THz frequencies, it is crucial to delve into studies to investigate further how THz propagation raises health concerns for the human body. Further investigation is needed to address the privacy and safety concerns in ultrafast communication, sensing, and localization networks. With the recent progress on improved beamforming gains and tunable beam steering technology, in conjunction with machine learning methods, the sensing and imaging data are more prone to being exposed to hacking, and see-through detection at the user end or over the network \cite{Huq8782882}. 

\end{enumerate}

\section{Conclusions}
\label{sec:conc}
In conclusion, non-destructive THz spectroscopic sensing with chemometrics is useful for material discrimination and has the potential to deal with real-life problems and applications. Employing machine learning methods can provide a powerful tool for both qualitative and quantitative analyses of THz spectral fingerprints. In this paper, we have successfully applied several relevant dimension reduction and classification techniques to identify solid and gaseous materials measured by transmission and absorption THz-TDS spectroscopy, respectively. We have demonstrated the feasibility of PCA, PLS, NMF, and t-SNE to reduce the high-dimensional THz data and extract its most prominent features. Furthermore, we employed SVM, LDA, KNN, and NB classifiers to determine the sample materials' quantitative determination or prediction. Our results confirm that PCA-GRNN and PLS-GRNN have superior performance among other machine learning models in identifying solid materials in the THz range. However, for the special case of gaseous materials, NMF-GRNN, comparable to PCA-GRNN, provides the best classification results over lower SNR values. Furthermore, we have presented a comprehensive overview of the machine learning-enabled technologies at the THz band. Subsequently, we have investigated the prospective applications, system design considerations, opportunities, and challenges surrounding JCAS systems. We have also outlined novel machine learning approaches that mitigate traditional learning techniques' challenges and provide robust, fast, and adaptive platforms for THz sensing and communications functionalities. We have also shed light on several research directions that are expected to guarantee optimal coverage and network design that will shape the deployment of next-generation JCAS systems, and we addressed the open health and privacy issues. Up to the authors' knowledge, this paper is the first article that presents a holistic overview of signal processing and machine learning techniques for efficient THz sensing and introduces a roadmap for several exciting future THz use cases.

% Generated by IEEEtran.bst, version: 1.14 (2015/08/26)

\section*{Biographies}
\footnotesize

\textbf{Sara Helal} received her B.S. degree in Electrical and Computer Engineering from Effat University, Saudi Arabia, in 2021. Her research interests are in the areas of statistical machine learning, signal and image processing, wireless communications, and mathematics. 

\textbf{Hadi Sarieddeen} (S'13-M'18) received his B.E. degree in Computer and Communications Engineering from Notre Dame University-Louaize, Lebanon, in 2013, and his Ph.D. degree in Electrical and Computer Engineering from the American University of Beirut (AUB), Lebanon, in 2018. He is currently a postdoctoral fellow at King Abdullah University of Science and Technology (KAUST), Thuwal, Saudi Arabia. His research interests are in the areas of wireless communications and signal processing for wireless communications.

\textbf{Hayssam Dahrouj} (S’02, M’11, SM’15) received his Ph.D. degree in electrical and computer engineering from the University of Toronto (UofT), Ontario, Canada in 2010. In July 2020, he joined the Center of Excellence for NEOM Research at KAUST as a senior research scientist. His main research interests include integrated space-air-ground communications, cross-layer optimization, cooperative networks, convex optimization, distributed algorithms, machine learning for wireless communications, and optical communications networks.

\textbf{Tareq Y. Al-Naffouri} (M'10-SM'18) received his Ph.D. degree in Electrical Engineering from Stanford University in 2004. He is currently a Professor at the Electrical and Computer Engineering department at KAUST. His research interests lie in the areas of sparse, adaptive, and statistical signal processing, localization, machine learning, and their applications.

\textbf{Mohamed-Slim Alouini} (S'94-M'98-SM'03-F'09) was born in Tunis, Tunisia. He received his Ph.D. degree in Electrical Engineering from Caltech, Pasadena, CA, in 1998. He served as a faculty member at the University of Minnesota, Minneapolis, then at Texas A\&M University at Qatar, Education City, Doha, Qatar before joining KAUST as a professor of Electrical Engineering in 2009. His current research interests include the modeling, design, and performance analysis of wireless communication systems.


\begin{thebibliography}{10}
\providecommand{\url}[1]{#1}
\csname url@samestyle\endcsname
\providecommand{\newblock}{\relax}
\providecommand{\bibinfo}[2]{#2}
\providecommand{\BIBentrySTDinterwordspacing}{\spaceskip=0pt\relax}
\providecommand{\BIBentryALTinterwordstretchfactor}{4}
\providecommand{\BIBentryALTinterwordspacing}{\spaceskip=\fontdimen2\font plus
\BIBentryALTinterwordstretchfactor\fontdimen3\font minus
  \fontdimen4\font\relax}
\providecommand{\BIBforeignlanguage}[2]{{%
\expandafter\ifx\csname l@#1\endcsname\relax
\typeout{** WARNING: IEEEtran.bst: No hyphenation pattern has been}%
\typeout{** loaded for the language `#1'. Using the pattern for}%
\typeout{** the default language instead.}%
\else
\language=\csname l@#1\endcsname
\fi
#2}}
\providecommand{\BIBdecl}{\relax}
\BIBdecl

\bibitem{bogue2018sensing}
R.~Bogue, ``Sensing with terahertz radiation: {A} review of recent progress,''
  \emph{Sensor Review}, 2018.

\bibitem{articleJepsen}
P.~Jepsen, D.~Cooke, and M.~Koch, ``Terahertz spectroscopy and imaging –
  modern techniques and applications,'' \emph{Laser \& Photonics Reviews},
  vol.~5, pp. 124--166, Jan. 2011.

\bibitem{sengupta2018terahertz}
K.~Sengupta, T.~Nagatsuma, and D.~M. Mittleman, ``Terahertz integrated
  electronic and hybrid electronic-photonic systems,'' \emph{Nature
  Electronics}, vol.~1, no.~12, p. 622, 2018.

\bibitem{8586961Li}
R.~{Li}, C.~{Li}, H.~{Li}, S.~{Wu}, and G.~{Fang}, ``Study of automatic
  detection of concealed targets in passive terahertz images for intelligent
  security screening,'' \emph{IEEE Transactions on Terahertz Science and
  Technology}, vol.~9, no.~2, pp. 165--176, Mar. 2019.

\bibitem{articleChaccour}
C.~Chaccour, M.~Soorki, W.~Saad, M.~Bennis, P.~Popovski, and m.~Debbah, ``Seven
  defining features of terahertz (thz) wireless systems: A fellowship of
  communication and sensing,'' \emph{IEEE Communications Surveys \& Tutorials},
  vol.~PP, pp. 1--1, 01 2022.

\bibitem{inproceedingsChaccour}
C.~Chaccour, R.~Amer, B.~Zhou, and W.~Saad, ``On the reliability of wireless
  virtual reality at terahertz (thz) frequencies,'' \emph{2019 10th IFIP
  International Conference on New Technologies, Mobility and Security (NTMS)},
  pp. 1--5, 2019.

\bibitem{AKYILDIZ201416Akyildiz}
I.~F. Akyildiz, J.~M. Jornet, and C.~Han, ``Terahertz band: {N}ext frontier for
  wireless communications,'' \emph{Physical Communication}, vol.~12, pp.
  16--32, 2014.

\bibitem{sarieddeen2020overview}
H.~Sarieddeen, M.-S. Alouini, and T.~Y. Al-Naffouri, ``An overview of signal
  processing techniques for terahertz communications,'' \emph{arXiv preprint
  arXiv:2005.13176}, 2020.

\bibitem{sarieddeen2019next}
H.~{Sarieddeen}, N.~{Saeed}, T.~Y. {Al-Naffouri}, and M.~{Alouini}, ``Next
  generation terahertz communications: {A} rendezvous of sensing, imaging, and
  localization,'' \emph{IEEE Communications Magazine}, vol.~58, no.~5, pp.
  69--75, 2020.

\bibitem{articleHsieh}
Y.-D. Hsieh, S.~Nakamura, D.~Ibrahim, T.~Minamikawa, Y.~Mizutani, H.~Yamamoto,
  T.~Iwata, F.~Hindle, and T.~Yasui, ``Dynamic terahertz spectroscopy of gas
  molecules mixed with unwanted aerosol under atmospheric pressure using
  fibre-based asynchronous-optical-sampling terahertz time-domain
  spectroscopy,'' \emph{Scientific Reports}, vol.~6, p. 28114, 06 2016.

\bibitem{9782674Chen}
H.~Chen, H.~Sarieddeen, T.~Ballal, H.~Wymeersch, M.-S. Alouini, and T.~Y.
  Al-Naffouri, ``A tutorial on terahertz-band localization for 6g communication
  systems,'' \emph{IEEE Communications Surveys \& Tutorials}, pp. 1--1, 2022.

\bibitem{faisal2019ultra}
A.~{Faisal}, H.~{Sarieddeen}, H.~{Dahrouj}, T.~Y. {Al-Naffouri}, and M.~S.
  {Alouini}, ``Ultramassive {MIMO} systems at terahertz bands: {P}rospects and
  challenges,'' \emph{{IEEE} Veh. Technol. Mag.}, vol.~15, no.~4, pp. 33--42,
  2020.

\bibitem{ruggiero2020invited}
M.~T. Ruggiero, ``Invited review: {M}odern methods for accurately simulating
  the terahertz spectra of solids,'' \emph{Journal of Infrared, Millimeter, and
  Terahertz Waves}, pp. 1--38, 2020.

\bibitem{qin2013detection}
J.~Qin, Y.~Ying, and L.~Xie, ``The detection of agricultural products and food
  using terahertz spectroscopy: {A} review,'' \emph{Applied Spectroscopy
  Reviews}, vol.~48, no.~6, pp. 439--457, 2013.

\bibitem{yin2016application}
M.~Yin, S.~Tang, and M.~Tong, ``The application of terahertz spectroscopy to
  liquid petrochemicals detection: {A} review,'' \emph{Applied Spectroscopy
  Reviews}, vol.~51, no.~5, pp. 379--396, 2016.

\bibitem{ELHADDAD201398}
J.~El~Haddad, B.~Bousquet, L.~Canioni, and P.~Mounaix, ``Review in terahertz
  spectral analysis,'' \emph{TrAC Trends in Analytical Chemistry}, vol.~44, pp.
  98--105, 2013.

\bibitem{gordon2017hitran2016}
I.~E. Gordon, L.~S. Rothman, C.~Hill, R.~V. Kochanov, Y.~Tan, P.~F. Bernath,
  M.~Birk, V.~Boudon, A.~Campargue, K.~Chance \emph{et~al.}, ``The {HITRAN2016}
  molecular spectroscopic database,'' \emph{Journal of Quantitative
  Spectroscopy and Radiative Transfer}, vol. 203, pp. 3--69, 2017.

\bibitem{heilweil2011thz}
E.~Heilweil and M.~Campbell, ``{THz} spectral database,'' 2011.

\bibitem{George2013}
D.~K. George and A.~G. Markelz, \emph{Terahertz Spectroscopy of Liquids and
  Biomolecules}.\hskip 1em plus 0.5em minus 0.4em\relax Berlin, Heidelberg:
  Springer Berlin Heidelberg, 2013, pp. 229--250.

\bibitem{hangyo2005terahertz}
M.~Hangyo, M.~Tani, and T.~Nagashima, ``Terahertz time-domain spectroscopy of
  solids: {A} review,'' \emph{International journal of infrared and millimeter
  waves}, vol.~26, no.~12, pp. 1661--1690, 2005.

\bibitem{HERRMANN2012107}
M.~Herrmann, F.~Platte, K.~Nalpantidis, R.~Beigang, and H.~M. Heise,
  ``Combination of kramers–kronig transform and time-domain methods for the
  determination of optical constants in thz spectroscopy,'' \emph{Vibrational
  Spectroscopy}, vol.~60, pp. 107--112, 2012.

\bibitem{Jornet5995306}
J.~M. Jornet and I.~F. Akyildiz, ``Channel modeling and capacity analysis for
  electromagnetic wireless nanonetworks in the terahertz band,'' \emph{IEEE
  Trans. Wireless Commun.}, vol.~10, no.~10, pp. 3211--3221, Oct. 2011.

\bibitem{Huang17}
Y.~Huang, P.~Sun, Z.~Zhang, and C.~Jin, ``Numerical method based on transfer
  function for eliminating water vapor noise from terahertz spectra,''
  \emph{Appl. Opt.}, vol.~56, no.~20, pp. 5698--5704, Jul 2017.

\bibitem{10.1117/12.527880}
B.~S. Ferguson, H.~Liu, S.~Hay, D.~Findlay, X.-C. Zhang, and D.~Abbott, ``In
  vitro osteosarcoma biosensing using {THz} time domain spectroscopy,'' in
  \emph{BioMEMS and Nanotechnology}, D.~V. Nicolau, U.~R. Muller, and J.~M.
  Dell, Eds., vol. 5275, International Society for Optics and Photonics.\hskip
  1em plus 0.5em minus 0.4em\relax SPIE, 2004, pp. 304 -- 316.

\bibitem{articleYan}
L.~Yan, C.~Liu, H.~Qu, W.~Liu, Y.~Zhang, J.~Yang, and L.~Zheng,
  ``Discrimination and measurements of three flavonols with similar structure
  using terahertz spectroscopy and chemometrics,'' \emph{Journal of Infrared,
  Millimeter, and Terahertz Waves}, vol.~39, 03 2018.

\bibitem{8977502Li}
C.~{Li}, B.~{Li}, and D.~{Ye}, ``Analysis and identification of rice
  adulteration using terahertz spectroscopy and pattern recognition
  algorithms,'' \emph{IEEE Access}, vol.~8, pp. 26\,839--26\,850, 2020.

\bibitem{articleCan}
C.~Cao, Z.~Zhang, X.~Zhao, and T.~Zhang, ``Terahertz spectroscopy and machine
  learning algorithm for non-destructive evaluation of protein conformation,''
  \emph{Optical and Quantum Electronics}, vol.~52, 04 2020.

\bibitem{190485Xudong}
X.~Sun, J.~Liu, K.~Zhu, J.~Hu, X.~Jiang, and Y.~Liu, ``Generalized regression
  neural network association with terahertz spectroscopy for quantitative
  analysis of benzoic acid additive in wheat flour,'' \emph{Royal Society Open
  Science}, vol.~6, no.~7, p. 190485, 2019.

\bibitem{026004Bowman}
T.~Bowman, T.~Chavez, K.~Khan, J.~Wu, A.~Chakraborty, N.~Rajaram, K.~Bailey,
  and M.~O. El-Shenawee, ``{Pulsed terahertz imaging of breast cancer in
  freshly excised murine tumors},'' \emph{Journal of Biomedical Optics},
  vol.~23, no.~2, pp. 1--13, 2018.

\bibitem{articleWendao}
W.~Xu, L.~Xie, Z.~Ye, W.~Gao, Y.~Yao, M.~Chen, J.~Qin, and Y.~Ying,
  ``Discrimination of transgenic rice containing the cry1ab protein using
  terahertz spectroscopy and chemometrics,'' \emph{Scientific reports}, vol.~5,
  p. 11115, 07 2015.

\bibitem{CHEN20151}
Z.~Chen, Z.~Zhang, R.~Zhu, Y.~Xiang, Y.~Yang, and P.~B. Harrington,
  ``Application of terahertz time-domain spectroscopy combined with
  chemometrics to quantitative analysis of imidacloprid in rice samples,''
  \emph{Journal of Quantitative Spectroscopy and Radiative Transfer}, vol. 167,
  pp. 1 -- 9, 2015.

\bibitem{GE2016286}
H.~Ge, Y.~Jiang, F.~Lian, Y.~Zhang, and S.~Xia, ``Quantitative determination of
  aflatoxin {B1} concentration in acetonitrile by chemometric methods using
  terahertz spectroscopy,'' \emph{Food Chemistry}, vol. 209, pp. 286--292,
  2016.

\bibitem{Zhan_2016}
H.~Zhan, K.~Zhao, H.~Zhao, Q.~Li, S.~Zhu, and L.~Xiao, ``The spectral analysis
  of fuel oils using terahertz radiation and chemometric methods,''
  \emph{Journal of Physics D: Applied Physics}, vol.~49, no.~39, p. 395101, sep
  2016.

\bibitem{LIU201886}
W.~Liu, C.~Liu, J.~Yu, Y.~Zhang, J.~Li, Y.~Chen, and L.~Zheng, ``Discrimination
  of geographical origin of extra virgin olive oils using terahertz
  spectroscopy combined with chemometrics,'' \emph{Food Chemistry}, vol. 251,
  pp. 86 -- 92, 2018.

\bibitem{articleLiu}
J.~Liu, ``Terahertz spectroscopy and chemometric tools for rapid identification
  of adulterated dairy product,'' \emph{Optical and Quantum Electronics},
  vol.~49, Jan. 2017.

\bibitem{045001Yury}
Y.~V. Kistenev, A.~V. Borisov, M.~A. Titarenko, O.~D. Baydik, and A.~V.
  Shapovalov, ``{Diagnosis of oral lichen planus from analysis of saliva
  samples using terahertz time-domain spectroscopy and chemometrics},''
  \emph{Journal of Biomedical Optics}, vol.~23, no.~4, pp. 1 -- 8, 2018.

\bibitem{sun2019generalized}
X.~Sun, J.~Liu, K.~Zhu, J.~Hu, X.~Jiang, and Y.~Liu, ``Generalized regression
  neural network association with terahertz spectroscopy for quantitative
  analysis of benzoic acid additive in wheat flour,'' \emph{Royal Society open
  science}, vol.~6, no.~7, p. 190485, 2019.

\bibitem{ye2020characterization}
D.~Ye, W.~Wang, H.~Zhou, H.~Fang, J.~Huang, Y.~Li, H.~Gong, and Z.~Li,
  ``Characterization of thermal barrier coatings microstructural features using
  terahertz spectroscopy,'' \emph{Surface and Coatings Technology}, vol. 394,
  p. 125836, 2020.

\bibitem{cui2020ultrafast}
J.~Cui, J.~Zhang, C.~Dong, D.~Liu, and X.~Huang, ``An ultrafast and high
  accuracy calculation method for gas radiation characteristics using
  artificial neural network,'' \emph{Infrared Physics \& Technology}, vol. 108,
  p. 103347, 2020.

\bibitem{8722599Petrov}
V.~{Petrov}, G.~{Fodor}, J.~{Kokkoniemi}, D.~{Moltchanov}, J.~{Lehtomaki},
  S.~{Andreev}, Y.~{Koucheryavy}, M.~{Juntti}, and M.~{Valkama}, ``On unified
  vehicular communications and radar sensing in millimeter-wave and low
  terahertz bands,'' \emph{IEEE Wireless Communications}, vol.~26, no.~3, pp.
  146--153, 2019.

\bibitem{9650789Saeed}
N.~Saeed, M.~H. Loukil, H.~Sarieddeen, T.~Y. Al-Naffouri, and M.-S. Alouini,
  ``Body-centric terahertz networks: Prospects and challenges,'' \emph{IEEE
  Transactions on Molecular, Biological and Multi-Scale Communications}, pp.
  1--1, 2021.

\bibitem{Lima9330512}
C.~De~Lima, D.~Belot, R.~Berkvens, A.~Bourdoux, D.~Dardari, M.~Guillaud,
  M.~Isomursu, E.-S. Lohan, Y.~Miao, A.~N. Barreto, M.~R.~K. Aziz,
  J.~Saloranta, T.~Sanguanpuak, H.~Sarieddeen, G.~Seco-Granados, J.~Suutala,
  T.~Svensson, M.~Valkama, B.~Van~Liempd, and H.~Wymeersch, ``Convergent
  communication, sensing and localization in 6g systems: An overview of
  technologies, opportunities and challenges,'' \emph{IEEE Access}, vol.~9, pp.
  26\,902--26\,925, 2021.

\bibitem{9634116Hadi}
H.~Sarieddeen, A.~Abdallah, M.~M. Mansour, M.-S. Alouini, and T.~Y.
  Al-Naffouri, ``Terahertz-band mimo-noma: Adaptive superposition coding and
  subspace detection,'' \emph{IEEE Open Journal of the Communications Society},
  vol.~2, pp. 2628--2644, 2021.

\bibitem{ryniec2012terahertz}
R.~Ryniec, P.~Zagrajek, and N.~Palka, ``Terahertz frequency domain spectroscopy
  identification system based on decision trees,'' \emph{Acta Physica
  Polonica-Series A General Physics}, vol. 122, no.~5, p. 891, 2012.

\bibitem{loukil2019terahertz}
M.~H. {Loukil}, H.~{Sarieddeen}, M.~S. {Alouini}, and T.~Y. {Al-Naffouri},
  ``Terahertz-band {MIMO} systems: {A}daptive transmission and blind parameter
  estimation,'' \emph{{IEEE} Commun. Lett.}, vol.~25, no.~2, pp. 641--645,
  2021.

\bibitem{ESNcoh2}
S.~P. {Chatzis} and Y.~{Demiris}, ``Echo state gaussian process,'' \emph{{IEEE}
  Trans. Neural Netw.}, vol.~22, pp. 1435--1445, Sep. 2011.

\bibitem{Parks21041186}
H.~Park and J.-H. Son, ``Machine learning techniques for thz imaging and
  time-domain spectroscopy,'' \emph{Sensors}, vol.~21, no.~4, 2021.

\bibitem{Huang9367503}
C.~Huang, Z.~Yang, G.~C. Alexandropoulos, K.~Xiong, L.~Wei, C.~Yuen, and
  Z.~Zhang, ``Hybrid beamforming for ris-empowered multi-hop terahertz
  communications: A drl-based method,'' in \emph{2020 IEEE Globecom Workshops
  (GC Wkshps}, 2020, pp. 1--6.

\bibitem{Boulogeorgos}
A.-A.~A. Boulogeorgos, E.~Yaqub, M.~di~Renzo, A.~Alexiou, R.~Desai, and
  R.~Klinkenberg, ``Machine learning: A catalyst for thz wireless networks,''
  \emph{Frontiers in Communications and Networks}, vol.~2, 2021.

\bibitem{Huq8782882}
K.~M.~S. Huq, S.~A. Busari, J.~Rodriguez, V.~Frascolla, W.~Bazzi, and D.~C.
  Sicker, ``Terahertz-enabled wireless system for beyond-5g ultra-fast
  networks: A brief survey,'' \emph{IEEE Network}, vol.~33, no.~4, pp. 89--95,
  2019.

\end{thebibliography}
\end{document}